\documentclass[lettersize,journal]{IEEEtran}
\usepackage{amsmath,amsfonts}
\usepackage{algorithmic}
\usepackage{algorithm}
\usepackage{array}
\usepackage[caption=false,font=normalsize,labelfont=sf,textfont=sf]{subfig}
\usepackage{textcomp}
\usepackage{stfloats}
\usepackage{url}
\usepackage{verbatim}
\usepackage{graphicx}
\usepackage{cite}

\usepackage{amsmath}
\usepackage{tabularx}
\usepackage{color}

\begin{document}

%\title{Survey on Energy-Aware Computing}
\title{Energy Concerns with HPC Systems and Applications}
\author{IEEE Publication Technology,~\IEEEmembership{Staff,~IEEE,}
\author{%%%% author names
    \IEEEauthorblockN{Roblex T. Nana \textsuperscript{+}}
    , \IEEEauthorblockN{Claude Tadonki \textsuperscript{+}}
    , \IEEEauthorblockN{Petr Dokladal \textsuperscript{++}
    , \IEEEauthorblockN{Youssef Mesri \textsuperscript{+++}}
    \\%%%% author affiliations
    \IEEEauthorblockA{\textit{\textsuperscript{+} Mines Paris - PSL, Centre de Recherche en Informatique (CRI), Fontainebleau, France}}\\
    \IEEEauthorblockA{\textit{\textsuperscript{++} Mines Paris - PSL, Centre de Morphologie Mathématique (CMM), Fontainebleau, France}}\\
    \IEEEauthorblockA{\textit{\textsuperscript{+++} Mines Paris - PSL, Centre de Mise en Forme de Matériaux (CEMEF), Sophia Antipolis, France}}\\
    %%%% corresponding author contact details
    \IEEEauthorblockA{Email : (roblex.nana\_tchakoute,claude.tadonki,petr.dokladal,youssef.mesri)}@minesparis.psl.eu}
}

        % <-this % stops a space
\thanks{This paper was produced by the IEEE Publication Technology Group. They are in Piscataway, NJ.}% <-this % stops a space
\thanks{Manuscript received April 19, 2021; revised August 16, 2021.}}



\maketitle

\begin{abstract}
For various reasons including those related to climate changes, {\it energy} has become a critical concern in all relevant activities and technical designs. For the specific case of computer activities, the problem is exacerbated with the emergence and pervasiveness of the so called {\it intelligent devices}. From the application side, we point out the special topic of {\it Artificial Intelligence}, who clearly needs an efficient computing support in order to succeed in its purpose of being an {\it ubiquitous assistant}.  There are mainly two contexts where {\it energy} is one of the top priority concerns: {\it embedded computing} and {\it supercomputing}. For the former, power consumption is critical because the amount of energy that is available for the devices is limited. For the latter, the heat dissipated is a serious source of failure and the financial cost related to energy is likely to be a significant part of the maintenance budget. On a single computer, the problem is commonly considered through the electrical power consumption. This paper, written in the form of a survey, we depict the landscape of energy concerns in computer activities, both from the hardware and the software standpoints.
\end{abstract}


\section{Introduction}
\label{sec:introduction}
\IEEEPARstart{M}{anufacturers} of high performance computing (HPC) systems are striving to provide more and more potential computing power by acting on the related hardware aspects  like: {\em number of cores, vectors units, 3-operands units, accelerators}, to name the main ones. Performance is a high priority in servers and supercomputers beside storage capacity. In order to leverage the potential power of HPC systems, efforts are made  reach better implementations through cutting-edge programming and code optimisation techniques \cite{tadHDR}. The reality is that these performance-guided activities do not explicitly consider the energy efficiency. Energy saving has become one of the main challenges for the new generations of servers and supercomputers. 

Nowadays, the design of HPC systems considers {\em power efficiency}: 21.1 MW for the 1.102 EFlop/s {\sc Frontier}, 29.9 MW for the 442 PFlop/s {\sc Fugaku}, and 2.94 MW for the 151.9 PFlop/s {\sc Lumi} to consider the top 3 machines of the most recent TOP500 list \cite{top500}. The associated electricity bill increasingly dominates the overall costs related all the activities of HPC systems.

The problem is generally formulated as the need to reach a good trade-off between {\em time-to-solution} and {\em energy-to-solution}. Different approaches have emerged to solve this problem, which can be summarised as follows: vendors work on power-eﬃcient processors and software developers on how to use them at the best \cite{article_1}. However, an effective solution is possible only by properly managing all layers of the system, from the software stack to the cooling system \cite{7092602}. Thus, we need power efficient software's as well as hardware integrated solutions and optimized devices.

The HPC market is growing significantly as the topic itself is becoming popular and the need for computing speed a genuine fact. The so-called "embedded HPC" is a new and emerging topic, which consists on the development and use of highly parallel micro-servers/embedded devices as mainframe computing systems \cite{CARDOSO201717}. These systems are increasingly used particularly in the field of artificial intelligence (AI) to support both {\em data collection} and {\em model inference}. The advantage of using embedded devices is their energy efficiency for a competitive computing performance compared to traditional CPUs. For machine/deep learning inference, a new generation of Coral Dev Bord micro-controllers can outperform traditional Intel Skylake server processors by more than 20x times on both time performance and energy efficiency\cite{devboard2020}. This illustrate the energy efficiency and the processing speed of embedded  systems over CPUs for some specifics applications. Thus, these low-power systems are widely considered as good candidates when energy is the central concern.

There are several contributions in the literature on energy in HPC and embedded systems. These works range from the definition of metrics \cite{book_1, Belady2008GREENGD} to the optimization of energy \cite{en16020890, Ramesh2012TechnicalRE, Claude2003} through the development of profiling and energy management tools\cite{9555945, Hackenberg2014HDEEMHD}. This work is carried out on the hardware and software side of the systems as well as on the algorithmic level, targeting different types of system {\it (embedded, CPU, GPU, FPGA and hybrid)}.

The focus on {\em energy} in the context of computer systems is also related to {\em carbon footprint}, which is a more general concern currently in the spotlight. Indeed, {\em energy} can be turned into {\em carbon emission} by multiplying it with the {\em carbon intensity} of the energy supply\cite{Patterson2022}. If the power consumption of most hardware components is well known or can be measured accurately, it not the case with carbon emission, which has to (roughly) estimated by specific means or using the aforementioned conversion.

In this paper, we survey a taxonomy of energy concerns in computers systems. For each type of system {\it (general purpose computers, accelerators, embedded systems/micro-controllers and modern supercomputers)}, we present stat-of-the-art (SOTA) architectures with a focus on power management tools. Another contribution of this work is a survey of SOTA energy/power optimization techniques with an emphasis on AI applications and a prospective analysis on all studied systems.

The remainder of this paper is structured as follows: Section \ref{sec:related} presents a review of existing surveys that address the topic of energy management in HPC and embedded systems. Section \ref{sec:metrics} is about a quantitative overview of energy/power aspect in computer, with a focus on the main energy/power and carbon footprint metrics. Section \ref{sec:hardware} show an overview of most recent energy aware hardware architecture for HPC and AI workloads. Section \ref{sec:management} discuss about a taxonomy of energy concern in embedded systems, accelerators, general purpose computers and modern supercomputers, with a special focus on energy management and optimization tools. Section \ref{sec:cooling} discuss about cooling system technologies from the energy consumption standpoint. Section \ref{sec:optimization} present, comments and discusses some energy/power optimization techniques from the literature. Section \ref{sec:ai_concern} present a literature review for energy concern of AI applications in computers system. Finally, Section \ref{sec:conclusion} concludes the paper.

\section{Related Review Works}\label{sec:related}

Beloglazov et al. \cite{article_2} discuss about the sources and issues of high power/energy consumption, and provide a taxonomy key aspects related to energy-efficient design of computing systems,  covering different levels including {\em  hardware, operating system, virtualization} and {\em data center}. The main aim of their taxonomy is to guide future design and development activities. 

A survey by Kocot et al. \cite{en16020890} investigates energy-aware scheduling methods used in modern HPC systems starting with the problem definition and then tackling various goals associated to this challenge, including a bi-objective approach that considers power and energy constraints. The work considers the standard types of HPC system (multicore CPU and GPU) together with related energy-saving mechanisms based on dynamic voltage/frequency scaling (DVFS), power capping, and other functionalities. The work uses a collection of carefully selected algorithms, classified by the programming paradigm (e.g. machine learning or fuzzy logic).

Czarnul et al. \cite{czarnul2019energy} provides a state of the art on energy-aware high-performance computing ({\em tools, techniques and environments}). They identify and classify the main approaches by {\em system/device types}, {\em optimization metrics}, and {\em energy/power control methods}. The work describes energy management tools (benchmarking, prediction, and simulation) and optimization approaches for standard devices (CPU/GPU/Hybrid) under various configurations (clusters, grids, and clouds). The authors point out the need for the unification of energy management interfaces for different architectures. Their conclusion states that we need to develop energy-aware methods for heterogeneous environments; indicates the optimization goals worth investigating based on minimizing the product of {\em energy} and {\em computing time}; and expresses the need for the validation of energy management tools.

An overview on energy-saving efforts is provided by Maiterth et al. \cite{maiterth2018energy}, where they focused on energy/power-aware job scheduling and resource allocation as a major step towards more efficient systems. The paper considers  nine large HPC centers located over three continents and the answers to eight questions from by their respective staff. Practical management procedures  including {\em power capping, job killing}, and  {\em virtual machines} are described. Moreover, the focus of the study is more  engineering oriented as it does not provide any formal or theoretical aspect related to energy-aware scheduling.

A survey by Chaudhry et al. \cite{chaudhry2015thermal} addresses thermal-aware scheduling and associated techniques for green data centers. Their study focuses on the thermal and cooling aspects of tasks scheduling, where a balanced heat distribution among the racks of the server is the main objective. They indicate some  metrics to evaluate thermal awareness in green data centers. In addition, they provide a thermal modeling together with effective solutions to prevent from hard-to-cool phenomena such as hot spots. They proposed two approaches: {\em reactive}, where the problem is fixed upon occurrence; and {\em proactive}, where the goal is to prevent the problem from occurring (e.g. using the thermal model of the server room followed by a proper tasks assignment on the compute nodes).

A technical report by Ramesh et al. \cite{Ramesh2012TechnicalRE} presents a taxonomy of power/energy concerns in embedded systems design. The proposed  taxonomy is derived from a systematic review of the literature, where a  categorization of the topics of interest is constructed. The authors considered a collection of 95 papers  related to energy management from ACM, IEEE Xplore, and SpringerLink databases. Their study focuses on energy dissipation and power optimization from the standpoint of hardware devices and that of support tools for energy profiling and optimization.

\begin{figure*}[h]
  \centering
  \includegraphics[scale=0.25]{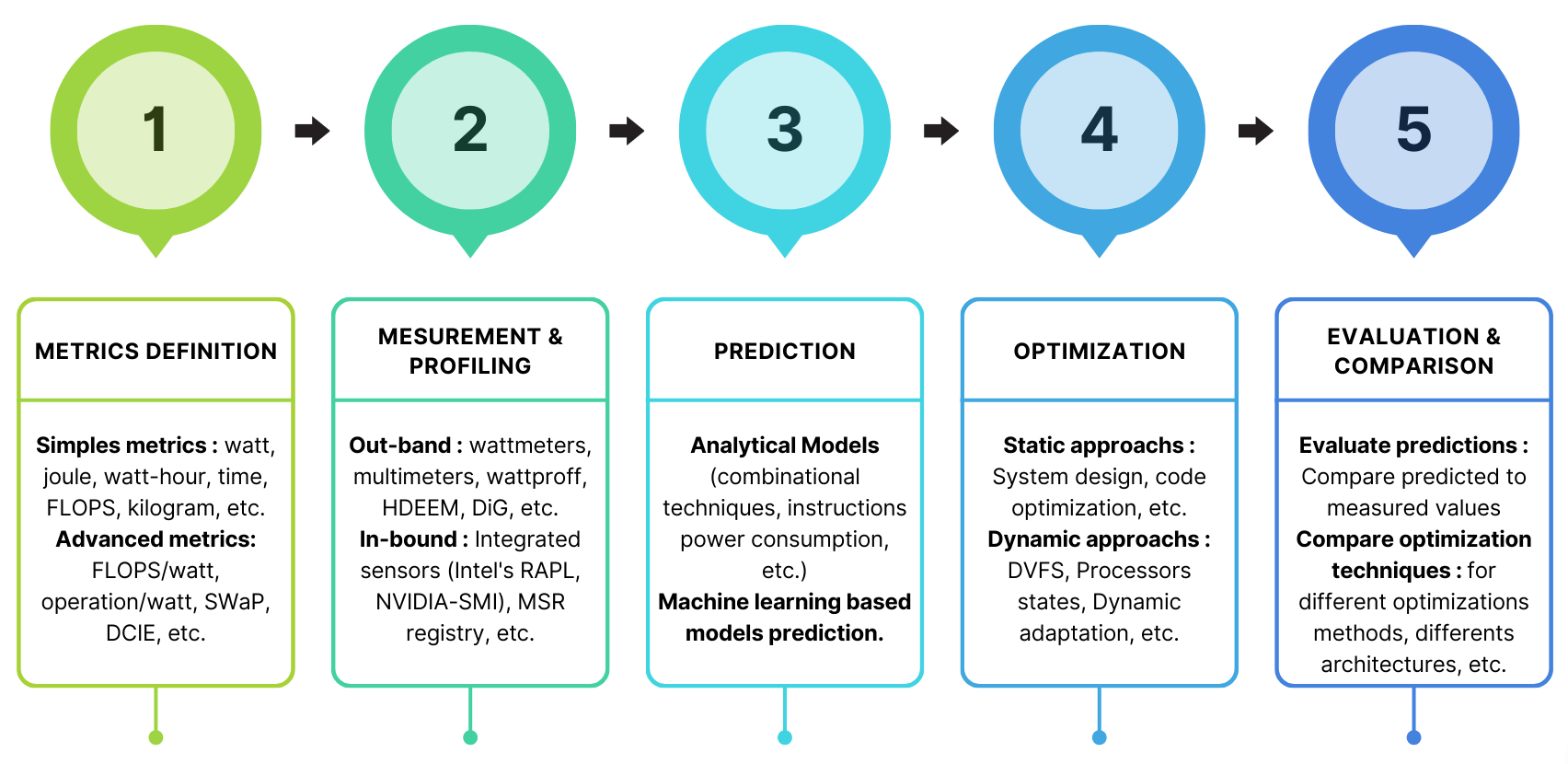}
  \caption{Taxonomy of power/energy management solutions with related approaches and tools.}
  \label{fig:taxonomy}
\end{figure*}

Many review works about energy concerns are generally specific to computer infrastructures (i.e., data-centers, embedded systems, supercomputer, etc.) for energy optimizations techniques, tools, and measurement. In this work, we survey the energy concern on HPC systems in a more general way considering all kind of approaches for energy/power management as presented in the taxonomy displayed in Figure \ref{fig:taxonomy}

\section{Overview of energy metrics in HPC}\label{sec:metrics}
The matter of appropriate energy and performance metrics has been investigated in several survey papers. However, since technology and associated features are evolving very rapidly, these studies lack some aspects that we present in this paper together with updated information related to news technologies. 

There are various approaches to power measurement and different types of outputs. We can classify theses measurement approaches into two groups: {\em Out-of-band} (e.g. power meters) and {\em in-band} (e.g. RAPL counters). Out-of-band measurement is the easiest approach to consider. It uses an external device to measure power consumption without a little to no interference in the  computational performance. In-band measurement requires some technical information about the target hardware and can access specific registers in a programmatic manner. Both types of measurement can be enhanced with an application-level profiling. However, it might be difficult to assess the type and detail of the measurements that are needed to obtain satisfactory insights from the energy profiling of the application. This is a major concern with the Out-of-band measurement, which uses an external device whose output data cannot be directly obtained within a program.

\subsection{Energy metrics in the supercomputer ecosystem}
Energy consumption is one of the major concerns when it comes to the deployment of large-scale HPC infrastructures. This must be taken into account at all levels {\em (from hardware to software tools)} and raises new scientific and operational challenges.

In the top500 ranking of June 2022\cite{top500}, the exascale performance (both theoretical and sustained) has been reached with {\sc Frontier} supercomputer \cite{Frontier2022}. The exascale was an important milestone in the HPC roadmap, and this level of potential performance is the current target of several high-end HPC infrastructures. The cost associated to the energy consumption by large-scale supercomputers is noticeable and the associated carbon footprint is becoming a serious concern. The Green500 \cite{green500} consider an evaluation of the {\em FLOPS per Watt} to rank supercomputers. Correlating the two metrics, we can state that the challenge is to increase the performance per energy consumed (FLOPS/Watt). Energy-efficient computing is a multi-dimensional problem, especially in the extreme-scale computing. The electricity consumption, thus the associated bill, includes the power due to machines operation and cooling system. A 2019 estimates ``A typical supercomputer consumes anywhere between 1 to 10 megawatts of power on average, which is equal to the electricity needs of almost 10,000 homes'' \cite{elec_houses}. For instance, the electricity bill paid by the RIKEN institute in 2020 for their (energy-efficient) {\em Fugaku} supercomputer was  nearly \$60 millions  \cite{elec:fugaku}. Table \ref{tab:elec_cost_hpc} give some illustrative data about the electricity bill of the top five supercomputers of the november 2022 top500 ranking \cite{top500}. We assume that the whole supercomputer is running continuously during 1 hour, thus we get the estimate electricity cost (in dollars, last column) by considering the cost per KWh that applies in the geographical location of the computing center. We considered the electricity prices per country on September 2022 \cite{price2022}.

\begin{table}[h]
    \centering
    \caption{Electricity cost per hour for the top five supercomputers.}
    \begin{tabular}{|l|l|r|r|r|}
         \hline
         {\bf Machine} &  {\bf Peak Perf.} &  {\bf Power(MW)} & {\bf \$/KWh} & {\bf Total(K\$)} \\
         \hline
         {\sc Frontier} & 1.685 EFLOPS & 21.1 & 0.150 & 3.165 \\
         \hline
          {\sc Fugaku} & 537.2 PFLOPS & 29.9 & 0.219 & 6.548 \\
         \hline
         {\sc LUMI} &  428.7 PFLOPS & 6.02 & 0.198 & 1.192 \\
         \hline
         {\sc Leonardo} &  255.7 PFLOPS & 5.61 & 0.561 & 3.147 \\
         \hline
         {\sc Summit} &  200.8 PFLOPS & 10.1 & 0.150 & 1.515 \\
          \hline
    \end{tabular}
    \label{tab:elec_cost_hpc}
\end{table}
\subsection{Specific energy-related metrics}
The so-called {\em thermal design power} (TDP), also called {\em thermal design point},  is the maximum amount of generated heat (by a computer chip or component) that the cooling system is designed to dissipate. The {\em power rating} (highest power input allowed) for a microprocessor is generally 1.5 times the TDP \cite{book_1}. The purpose of the TDP is to provide system designers with a power target so as to guide the selection of a convenient thermal solution. Under a steady workload, the TDP is the maximum power consumption of the processors. However, during the turbo mechanism or certain types of  workload such as vectors instructions, it can sometimes exceed the maximum TDP.

The so-called {\em Average CPU Power} (ACP), a concept defined by AMD for Opteron processors, is the average dissipated power of a processor while running a defined set of benchmarks  (Transaction Processing Performance Council (TPC Benchmark*-C), SPECcpu*2006, SPECjbb*2005, and STREAM) \cite{acp2011}. AMD indicated that these measurements to determine the ACP value are not to be considered for every processor, but only for some particular ones selected by its manufacturing units \cite{acp2022}.

From thermal standpoint, the processor TDP specification is a critical value because any thermal solution should dissipate at the level of that rated indication. Intel and AMD both agree on this point. If a given processor design is based on the ACP, then it might be undersized and out of its thermal specifications. For servers the main concern is not on how much power a specific component dissipates, but instead the power of the entire server when running a given workload. The corresponding measurement can be easily made following and {\em out-to-band} approach with a power meter on the input cord(s).

SWaP (Space, Wattage and Performance) \cite{SWaP2013} is an objective three-dimensional metric that provides a more comprehensive and realistic way to evaluate servers. It is calculated considering {\em performance, power} as indicated by equation \ref{eq_SWaP} that follows:  
\begin{equation}
\label{eq_SWaP}
    SWaP = \frac{Performance}{Space * Power}
\end{equation}
where {\em Performance and Power} is measured by any convenient benchmarks, and {\em Space} is related to the size of the computer.

The so-called {\em Power usage effectiveness} (PUE)\cite{Belady2008GREENGD} is a metric used to determine the energy efficiency of a data center. It is determined by dividing the total amount of incoming power by the consumed power as expressed by formula \ref{eq_PUE}.
\begin{footnotesize}
\begin{equation}
\label{eq_PUE}
    PUE = \frac{Total{\_}Facility{\_}Energy}{IT{\_}Energy} = 1+\frac{Non{\_}IT{\_}Energy}{IT{\_}Energy}
\end{equation}
\end{footnotesize}
According to the \textit{"Uptime Institute Annual Global Data Center Survey 2021"} \cite{uptime2021}, PUE and power consumption are among the top tracked sustainability metrics. But in 2022, key findings reported in the Uptime Institute Global Data Center Survey 2022 \cite{uptime2022} indicated the requirement of additional metrics to supplement PUE for future efficiency gains, which should focus on IT power. A similar benchmarking standard considered by the Green Grid is DCiE ({\em Data Center Infrastructure Efficiency}), which is just the inverse of PUE. Both metrics apply to a more global level, thus they do not capture the consumption specific to the computing activities. Indeed, information technology (IT) equipment include computing units and all associated peripherals. Nevertheless, having such a macroscopic information makes sense as all considered facilities are related to the computing activities.

A survey by Jin et al. \cite{Jin2016GreenDC} presents the state-of-the-art on green data center techniques including {\em energy efficiency, resource management, thermal control} and {\em green metrics}, with a detailed comparison among them and key challenges for future research. 

\subsection{From energy to Carbon footprint}\label{sec:co2}
Climate change currently stands as a critical concern because of its significant impact on ecosystems and livelihoods across the world. It's a clear fact that carbon dioxide emissions are the primary driver of global climate change. According to recent estimates, the total $CO_{2}$ emissions of the information and communications technology (ICT) sector account for around 2.1\%–3.9\% of global $CO_{2}$ emissions\cite{FREITAG2021100340}. Therefore, estimating and reducing the carbon footprint in ICT is worth all related efforts.

The typical way for carbon footprint estimate of IT infrastructure activities is to derivative it from power consumption. The paper by Patterson et al \cite{Patterson2022} provides a valuable study of the carbon footprint of computing workloads. They stated that  {\em $CO_{2}$ equivalent emissions $(CO_{2}e)$} accounts for {\em carbon dioxide} and all the other greenhouse gasses as well like {\em methane and nitrous oxide} for instance. This equivalent emission can be calculated from the {\em electric power} by multiplying it with the {\em carbon intensity} of the energy supply as expressed through formula (\ref{co2_foot}) \cite{Patterson2022}:
\begin{equation}
\label{co2_foot}
    CO_{2}e = Wh * (CO_{2}e \ per \ Wh)
\end{equation}

{\em Carbon intensity ($CO_{2}e$ per Wh)} is the amount of carbon dioxide $(CO_{2}e)$ that is released to produce a watt-hour of electricity. The average data-center carbon emissions in 2020 was 0.429 $tCO_{2}e$ (ton of carbon dioxide equivalent emissions) per MWh (Megawatt hour), but the gross $tCO_{2}e$ per MWh can be 5x lower in some specific data-centers \cite{Patterson2022}.
Table \ref{tab:co2_cost_hpc} is provided as an illustration of the carbon footprint for top ranked supercomputers. We used formula  (\ref{co2_foot}) and an estimation of the {\em carbon intensity} from 2022 data \cite{elec_carbon}. We clearly see that the floating-point performance and the necessary $(CO_{2}e)$ are not directly correlated, the hardware profile of the machines is a key factor.

\begin{table}[h] 
    \centering
    \caption{CO$_2$ per hour for the top five supercomputers.}
    \begin{tabular}{|l|l|r|r|r|}
         \hline
         {\bf Machine} &  {\bf Peak Perf.} &  {\bf Power} & {\tiny \bf Kg(CO$_2$)/KWh} & {\bf \footnotesize Kg(CO$_2$)} \\
         \hline
         {\sc Frontier} & 1.685 EFLOPS & 21.1MW & 0.379 & 7 997 \\
         \hline
          {\sc Fugaku} & 537.2 PFLOPS & 29.9MW & 0.479 & 14 322 \\
         \hline
         {\sc LUMI} &  428.7 PFLOPS & 6.02MW & 0.132 & 795 \\
         \hline
         {\sc Leonardo} &  255.7 PFLOPS & 5.61MW & 0.372 & 2 087 \\
         \hline
         {\sc Summit} &  200.8 PFLOPS & 10.1MW & 0.379 & 3 828 \\
          \hline
    \end{tabular}
    \label{tab:co2_cost_hpc}
\end{table}

\section{Common devices from the energy standpoint}\label{sec:hardware}

\subsection{Accelerators}

\subsubsection{GPU}
GPUs are specialized devices designed for efficient graphics rendering and image processing. Their parallel structure makes them more efficient than {\em traditional CPUs} for algorithms that process large blocks of data in parallel. Nowadays, GPUs stand as the reference accelerator in the HPC landscape. In addition to being now designed as general purpose units, GPUs have a top consideration when it comes to energy efficient in HPC \cite{Ikram2018}. From the absolute standpoint, modern GPUs consume a significant amount of power (from 50-600W or even more). However, because of their noteworthy processing speed, they show better performance-per-watt than standard CPUs for specific workloads.

AMD Instinct MI250X was ranked world’s fastest HPC accelerator in 2022\cite{amd2021}. This GPU has a (double-precision) peak performance of 47.9 TFLOPS and a peak power between 500 and 560 TDP. A combination with cutting-edge processors yield very powerful HPC systems, like the (AMD EPYC CPU, AMD Instinct MI250X) CPU-GPU pairing of the {\sc Frontier} (Exascale) supercomputer and other top ranked machines from the top500 list of November 2022\cite{top500}. NVIDIA H100 Tensor Core GPU \cite{nvidiaGPU} is the response from NVIDIA about this innovation from AMD as competitive material with 700W TDP in maximum configuration. Intel Launched {\em Intel Data Center GPU Max Series} project in 2022 with {\sc Ponte Vecchio} as the first competitive product for data center GPU market \cite{IntelGPU2022} with 600W TDP. A comparative view in terms of {\em flops performance} and {\em power supply} of the major accelerators is provided in table \ref{tab:tab_accelerator}.

\subsubsection{TPU}

With the high computing power required for cutting-edge AI, domain-specific architectures for Neural Network computations have emerged, like the {\em Tensor Processing Unit (TPU)}, a Deep Neural Network (DNN) accelerator from Google. 
An individual Edge TPU can perform 4 trillion operations per second (4 TFLOPS) with only 2 watts of power. The latest TPU ({\em version 4}) has an average TDP of 192W. For illustration, the Edge TPU can execute state-of-the-art mobile vision models such as MobileNet V2 at almost 400 frames per second in a power efficient manner \cite{tpu2023}. Pandey P. et al \cite{9586224} parameterized the extreme hardware under-utilization in a TPU systolic array and proposed UPTPU: an intelligent  data-flow adaptive power-gating paradigm that yields a improvement of the TPU energy efficiency by factor 3.5 to 6.5 on different input batch sizes. This ultra low power devices is nowadays integrated as an accelerator into microcontrollers (single-board unit), as we can see with {\em Coral Dev Board} for instance.

\subsubsection{FPGA}
Field Programmable Gate Arrays (FPGAs) are integrated circuits with ability to be reconfigured to implement a specific processing at the hardware level. Initially applicable to very specific domains, FPGAs has extended so as to now stand as a important components of servers and supercomputers, as well as edge computing systems\cite{fpga2022}. However, their energy efficiency is still an important concern, with no easy or standard ways for hardware/software power management. 

Hosseinabady and Nunez-Yanez \cite{Hosseinabady2018DynamicEM} investigate the use of FPGAs in an embedded system for energy saving. They study the energy efficiency of a hybrid FPGA-CPU device that can switch between hardware and software on periodic tasks. In addition, they successfully applied the voltage and frequency scaling (VFS) to reduce the energy consumption. Moreover, they showed that in some cases, if the task’s period is higher than a specific threshold a reduction of the energy consumption cannot be obtained on the FPGA, hence the effectiveness of a software support for energy saving. Experimental results show up to 48\% energy reduction by applying the proposed techniques at runtime on thirteen individual tasks. As previously said, the major accelerator in the HPC landscape remains the GPU, however the FPGA is becoming a serious candidate.

\begin{center}
\begin{table*}
    \caption{SOTA accelerators systems characteristics}
    \begin{tabularx}{\textwidth}{|c|c|X|X|l|X|X|} 
        \hline
        {\bf Name} & {\bf memory (GB)} & {\bf core frequency (GHz)} & {\bf TDP (w)} & {\bf peak TOPS} & {\bf peak TFLOPS (fp32)} & {\bf performance/watt (INT8)} \\ 
        \hline
        Tesla A100 SXM4 & 80 & 1.41 & 400 & 312(bf16)/624(int8) & 19.5 & 1.56 TOPS/W \\ 
        \hline
        Tesla H100 SXM5 & 80 & 1.98 & 700 & 1000(bf16)/2000(int8) & 60 & 3.33 TOPS/W \\ 
        \hline
        AMD Instinct MI250X & 128 & 1.7 & 560 & 383 (bf16 or int8) & 95.7 & 0.68 TOPS/W \\ 
        \hline
        Intel Ponte Vecchio & 128 & 1.6 & 600 & 720(bf16)/1440(int8) & 45 & 2.40 TOPS/W \\ 
        \hline
        Google TPU v4 & 32 & 1.05 & 192 (idle) & 275 (bf16 or int8) & / & 1.43 TOPS/W \\ 
        \hline
    \end{tabularx}
    \label{tab:tab_accelerator}
\end{table*}
\end{center}

\subsection{Embedded systems: Microcontrollers}
Due to their small size and single-chip configuration (thus at the expense of processing power, memory and storage), micro-controllers have a little energy consumption while keeping a certain level of computing efficiency. Much more energy is required to power a GPUs and standard CPUs, which yields some limitations and constraints in their usage. Micro-controllers are typically not wired into main power, they instead rely on batteries or residual energy. For example, a micro-controller can run on a single coin battery for weeks or even months. However, having a low power system does not yield lower energy consumption by itself. Indeed, it is important to optimize the software, not just in terms of functionality or processing efficiency, but also with respect to energy efficiency. We now describe some of the major devices of embedded computing.

\subsubsection{Arduino}
When it comes to microcontrollers and embedded systems, one of the first candidate that pops is {\em Arduino}. Arduino is an open-source electronics platform based on easy-to-use hardware and software for ultra low power chips, which consumes less than a single watt of nominal power. The Arduino Portenta H7 \cite{Arduino} is currently the most powerful IoT Cloud compatible boards of the Arduino series. Arduino can be used to connect devices, visualize data, control and share projects online. Beginners and advanced users can meet their   specific needs from its wide range of features and possibilities. However, even with Portenta H7, Arduino is not powerful enough to handle HPC workloads in the context of embedded systems as compared with others microcontrollers (see Table \ref{tab:tab_Embedded_systems}).

\subsubsection{Raspberry Pi}
While Arduino is an electronic board with a simple microcontroller, Raspberry Pi is a full-fledged computer. Unlike Arduino, Raspberry Pi has its own operating system, thus it can carry out more complex operations (e.g. {\em robot control} and  {\em weather monitoring}, to name these two). 
The {\em Raspberry Pi 4 Model B} \cite{RaspberryPi} is the latest model of the Raspberry Pi microcontrollers series. It offers a noteworthy increase in processing speed, multimedia performance, memory and connectivity over the previous generation (Raspberry Pi 3 Model B+), while keeping full compatibility with earlier versions and same level of power consumption. The Model B offers a level of performance comparable to that of entry-level x86 PC systems but with the advantage of energy efficiency as it has a maximum nominal power of 10W.

\subsubsection{Intel NCS2}
Intel Neural Compute Stick 2 (Intel NCS2)\cite{ncs22022} is a {\em plug-and-play} Development Kit for AI Inference. NCS2 is based on an Intel Vision Processing Unit(VPU) chip named  Movidius X. Movidius provides its Neural Compute Stick (i.e. {\em Fathom}) to bring a basic-level deep learning capabilities into embedded devices. It can be used to develop, fine-tune, and deploy convolutional neural networks (CNNs) on low-power applications that require real-time inference. It supports  heterogeneous execution across computer vision accelerators (CPU, GPU, VPU, and FPGA) using a unified API. Its so-called Vision Processing Unit (VPU) includes vision accelerators, a Neural Compute Engine, imaging accelerators, and 16 SHAVE vector processors paired with a CPU in one heterogeneous package. The combination of the aforementioned units provides a total of up to 4 TFLOPS with 1.5W of power \cite{ncs22017}. However, Intel is discontinuing this product and its technical support will continue until June 30, 2023, while warranty support will continue until June 30, 2024\cite{ncs22022}. 

\subsubsection{Nvidia Jetson}
The Jetson Nano Developer Kit \cite{JetsonNano} is the most popular board from Nvidia Jetson series. It delivers a noteworthy processing capability to efficiently support high-performance AI at low power and cost. The developer kit can be powered by micro-USB and comes with extensive I/Os. This makes it simple for developers to connect a diverse set of new sensors to enable a variety of applications at a little power of 5 watts. The Jetson AGX Orin Developer Kit \cite{JetsonAGX} is currently the most powerful board from this series, with up to 275 TOPS for running the NVIDIA AI software stack. It enables to create advanced robotics and edge AI applications. But this performance incurs a higher cost with currently more than 2000\$ for 60W TDP.

\subsubsection{Coral Dev Board}
The Coral Dev Board\cite{coral2020} is a single-board computer with a removable system-on-module (SOM) that contains eMMC, SOC, wireless radios, and Google’s Edge TPU. It’s perfect for IoT devices and other embedded systems that demand fast on-device ML inferencing. Coral dev is also the most efficient out of all the microcontrollers we have found. With on board TPU, it is capable of performing 4 tera-operations per second (TOPS), using 0.5 watts for each TOPS (2 TOPS per watt)\cite{devboard2020}. The USB version can be connect to any system running Debian Linux (including Raspberry Pi), macOS, or Windows 10. Coral Dev is very fast but, with bad tech support, faulty units and seems like a very common problem.

\begin{center}
\begin{table*}
    \caption{Characteristics of selected state of the art embedded systems }
    \begin{tabularx}{\textwidth}{|c|c|X|X|X|c|c|} 
        \hline
        {\bf Name} & {\bf memory(GB)} & {\bf core frequency (GHz)} & {\bf nominal power (w)} & {\bf peak TOPS} & {\bf peak TFLOPS (fp32)} & {\bf performance/watt} \\ 
        \hline
        Raspberry Pi 4B & 8 & 1.5 & 10 & / & 0.135 & 2.02 GFLOPS/W \\ 
        \hline
        Jetson Nano & 4 & 1.43 & 10 & 0.472(int8) & 0.236 & 0.047 TOPS/W \\ 
        \hline
        Jetson AGX Orin & 64 & 2.0 & 60 & 275(int8) & 5.3 & 4.58 TOPS/W \\ 
        \hline
        Arduino Portenta H7 & 0.008 & 0.48 & 1.15 & / & / & / \\ 
        \hline
        The Coral Dev Board & 4 & 1.5 & 0.65 & 4(int8) & / & 2 TOPS/W \\ 
        \hline
        Intel NCS2 & 8 & 0.7 & 1.5 & 4(int8) & / & 2.66 TOPS/W \\ 
        \hline
    \end{tabularx}
    \label{tab:tab_Embedded_systems}
\end{table*}
\end{center}

\subsection{General Purpose Processors}
\subsubsection{x86 based processors}
{\em x86} is a family of CISC instruction set architectures, initially developed by Intel from Intel 8086 microprocessor and its 8088 variant. It was introduced in 1978 as a fully 16-bit extension of Intel's 8-bit 8080 microprocessor, with memory segmentation as a solution for addressing more memory than can be covered by a plain 16-bit address. Embedded systems and general-purpose computers used x86 chips before the IBM Personal Computer in 1981.
Nowadays, most desktop, workstation, laptop and server computers are based on the x86 architecture family, while mobile categories such as smartphones or tablets are dominated by ARM.  The fastest supercomputer in the TOP500 list for November 2022 ({\sc Frontier}) is built with {\em AMD Epyc} CPUs that are based on the x86 ISA. The market of CPUs in the HPC landscape and data centers is still dominated today by x86 CPUs.

AMD claim that its EPYC processors power the most energy-efficient x86 servers, delivering exceptional performance with lower energy consumption \cite{amd_vs_intel_cpu}. AMD EPYC 9654 servers shall use up to 29\% less annual power  than Intel Xeon Platinum 8490H servers at the same performance, while helping reduce capital expenditure up to 46\% \cite{amd_vs_intel_cpu}. Note that these two CPU models require respectively 350W and 360W for Intel and AMD CPUs respectively.

\subsubsection{ARM based processors}
Energy saving has become one of the main challenges for new generation servers and supercomputers. Many manufacturers of HPC systems consider low-power ARM components that are also present today in the vast majority of embedded or mobile systems. Indeed, the particularity of ARM components is their low energy consumption with a competitive processing performance as Intel and AMD x86 architectures.
Several international collaborative projects like the Japanese Post-K, the European Mont-Blanc, or the UK’s GW4/EPSRC, announced the adoption of ARM technology for their high-performance computing (HPC) systems \cite{Mantovani2020PerformanceAE}. On November 2018, for the first time, an ARM-based system was listed in the Top500 ranking. It was the Astra\cite{astra} supercomputer powered by Marvell’s ThunderX2 ARM CPU and hosted at the Sandia National Laboratories (USA).

\begin{center}
\begin{table*}
    \caption{SOTA General purpose computers characteristics}
    \begin{tabularx}{\textwidth}{|c|c|X|X|X|X|c|} 
        \hline
        {\bf Name} & {\bf RAM (GB)} & {\bf core frequency (GHz)} & {\bf TDP (w)} & {\bf peak TOPS} & {\bf peak TFLOPS (fp32)} & {\bf performance/watt (INT8)} \\ 
        \hline
        Intel Platinum 8490H & 4000 & 1.9 & 350 & / & 3 648 & 10.42 GFLOPS/W \\ 
        \hline
        AMD EPYC 9654 & 6000 & 2.4 & 360 & 7763(int8) & 3 686 & 10.23 GFLOPS/W \\ 
        \hline
        Fujutsu A64FX & 32 & 2.6 & 150 & 3.4(int8) & 3 400 & 22.66 GFLOPS/W \\ 
        \hline
        Marvell ThunderX2 & 512 & 2.2 & 180 & / & 563 & 3.12 GFLOPS/W \\ 
        \hline
    \end{tabularx}
    \label{tab:tab_CPU_system}
\end{table*}
\end{center}

\section{Energy management tools}\label{sec:management}

\subsection{Energy tools for GPUs}
{\em NVIDIA-SMI (NVIDIA System Management Interface)} \cite{nvidia2016} is a command line utility for the management and monitoring of NVIDIA GPU devices that is based on the NVIDIA Management Library (NVML). The tool can be used to set the power range (max and min, in Watt) of the execution of a given application. Its GPU Operation Mode (GOM) allows to reduce the power usage and optimize the GPU throughput by disabling some features accordingly. It also implements a power scaling algorithm to dynamically reduce the clock frequency when the GPU is consuming too much power.

%{\color{blue}
{\em ROCM-SMI (ROCm System Management Interface)} \cite{ROCm-SMI} is part of ROCm (Radeon Open Compute), which is an open-source stack for GPU computation that offers several programming models like HIP, OpenMP, Message Passing Interface (MPI), and OpenCL. ROCM-SMI provides functionalities for clock, memory, power and temperature management of your ROCm enabled system. Similarly to NVIDIA-SMI, ROCM-SMI can be used for GPUs energy profiling, monitoring and optimization. The tool implements dynamics optimization and power capping techniques to reduce the energy consumption. ROCM-SMI provides a (Linux) command line mode and a python API for programming considerations.
%}

\begin{center}
\begin{table*}
    \caption{SOTA accelerators energy/power management tools}
    \begin{tabularx}{\textwidth}{|c|c|X|X|X|} 
        \hline
        {\bf Name} & {\bf type} & {\bf objective} & {\bf techniques} & {\bf portability} \\ 
        \hline
        NVML(NVIDIA-SMI)\cite{nvidia2016} & Software & device management & Dynamic Power Management, Power capping, measurement, etc. & Linux with Nvidia GPU devices; not tested it on Windows \\ 
        \hline
        ROCM-SMI\cite{ROCm-SMI} & Software & device management & Dynamic Power Management, Power capping, measurement, etc. & Linux with AMD ROCm devices; not tested it on Windows \\ 
        \hline
    \end{tabularx}
    \label{tab:tab_accelerator_tools}
\end{table*}
\end{center}

\subsection{Energy tools for CPUs}
- Intel RAPL\cite{RAPL2022}(Running Average Power Limit Interface) is an interface for reporting the (accumulated) energy consumption of various system-on-chip (SoC). The RAPL's energy reporting feature has been available on many generations of Intel SoC products. Intel processors utilize this energy information for internal SoC management purposes such as the control of power limits in association with the Turbo Boost power limit settings. Energy information from the RAPL interface gets updated every ~1 ms, which is several orders of magnitude slower than what physical side channel probing could achieve. RAPL measurements ignore a large part of the power consumption of servers because they focus on CPU and RAM. Some experiments on Intel processor from Grid5000 \cite{grid5000} show that it just represent 42\% of the overall servers consumption \cite{anne2021}.

- AMD RAPL counters : Concerning Zen architecture, AMD replaced APM (Application Power Management) with RAPL. The implementation is similar to the corresponding Intel's RAPL, but uses different control registers. While Intel typically provides multiple domains and the option to limit power consumption over various time frames, AMD only considers registers for memory reads and core power consumption. However, the latter is available with a per-core spatial resolution, while a per-package applies for Intel's core domain. Schöne et al. \cite{Schne2021EnergyEA} highlighted various energy efficiency aspects of the AMD Zen 2 micro-architecture in order to facilitate system comprehension and optimization. Key findings include qualitative and quantitative descriptions regarding {\em core frequency transition delays}, {\em workload-based frequency limitations}, and {\em effects of I/O die P-states on memory performance}. The authors made a comparative study with some high-end Intel architectures (i.e., Cascade Lake, Skylake, Haswell) for power efficiency and provided details on power measurements accuracy on both architectures. The work shows that AMD RAPL is unsuitable to optimize the overall energy consumption. Their approach failed on reflecting the influence of the operands, which can also be seen as a benefit when it comes to side-channel attacks that are based on power measurement.

- For ThunderX2\cite{tx2} chips, there is no RAPL counters but there are other {\em harware specific} on-chip sensors. These sensors are not yet supported by common libraries like PAPI\cite{Browne1999PAPIAP}, perf-tools\cite{perftools} for instance. However, Marvell\cite{tx2} has provided an tool named {\em tx2mon} \cite{tx2mon}, which is based on the Linux kernel driver {\em tx2mon\_kmod} to provide access to specific system data and allow to configure the way to measure energy.

- Model-Specific Register (MSR) is any of the various control registers in the x86 architecture used for debugging, program execution tracing, computer performance monitoring, and toggling certain CPU features.

- ACPI (Advanced Configuration and Power Interface)\cite{acpi} is an open standard that the operating system can use to discover and configure the components of the computer, to perform power management, auto configuration, and status monitoring. ACPI defines the performance states, designated by P-States. P-States correspond to different performance levels that apply while the processor is actively executing instructions according to energy saving and performance trade-off scenarios. Each system manufacturer decides its way to implement this specification standard to save energy in the system. For example, Intel CPUs, regarding Haswell architecture, provides voltage regulators per core, thus each core has its own P-State.

- Device Tree (DT)\cite{devicetree}: While ACPI was historically created for x86 platforms, the ARM ecosystem developed "Device Tree" (DT) to describe the same information for ARM-based devices. Thus, ACPI and DT overlap in that they both provide mechanisms for enumerating devices and attaching additional configuration data to devices (which can be used by higher layers of software). The biggest difference between DT and ACPI is that DT is effectively a structured mechanism for passing arbitrary data, while ACPI provides standardised data.

- PAPI (Performance API) \cite{Browne1999PAPIAP} library is a platform independent tool which provides developers with an interface and methodology for gathering performance-related hardware data. The basic principle is to allow developers to see the relation between the software performance and corresponding processor events. McCraw et al. \cite{PAPIenergy} extended PAPI to measure and report energy and power values even on complex architectures.

- Intel Power Gadget \cite{powergadget} is one of the most easy-to-use energy profiler. It provides a graphical user interface with a few plots showing CPU and DRAM utilisation (\%), cores frequency (GHz), temperature (ºC), and power consumption (W). The total energy consumption of the CPU and DRAM written into files (i.e. {\em Log files}). When installing Intel Power Gadget, its command-line interface (named PowerLog) is also installed.

- Powerstat \cite{powerstat} is an easy-to-use tool to measure energy consumption on Linux. Intel Power Gadget and PowerLog are not compatible for Linux system, so Powerstat was developed similarly to the previous tools. Powerstat is just another wrapper around an Intel library RAPL. However, it provides a simple interface for a command-line usage. 

- PowerTOP\cite{powertop} is a Linux tool used to diagnose issues with power consumption and power management. In addition to being a diagnostic tool, PowerTOP also has an interactive mode that can be used to handle various power management settings in case the direct mode is restricted by the OS. Its main advantage is the ability to estimate the energy consumption of the considered machine. It provides an interactive mode to fine-tune power management settings in Linux system.

- Perf tools\cite{perftools}: A very quick and easy way to obtain the energy consumption of a program in a Linux environment, is through Perf. It is a command-line tool that offers a wrapper to Intel’s RAPL. It facilitates the collection of energy measurements of the components of a computer and associated devices.

- Another quick way of getting energy and power measurements for Intel processors is through Likwid\cite{likwid2010}. Likwid uses the RAPL interface, developed by Intel, to fetch energy and power measurements from different types of CPU. Compared to Perf, Likwid does not offer an option to run a given test several times. However, it provides {\em power} estimation in addition to energy measurement.  Moreover, Likwid offers other options such as thread's temperature monitoring.

- PyJoules\cite{pyjoules} is a software toolkit written in Python to measure the energy footprint of a given host machine. It monitors the energy consumed by a specific device of the host machine. It works ionly with intel CPUs, RAM (for intel server architectures), intel integrated GPUs  and nvidia GPUs. 

\begin{center}
\begin{table*}
    \caption{SOTA General purpose computers tools for energy/power management}
    \begin{tabularx}{\textwidth}{|c|c|X|X|X|} 
        \hline
        {\bf Name} & {\bf type} & {\bf objective} & {\bf techniques} & {\bf portability} \\ 
        \hline
        RAPL counters\cite{RAPL2022} & hardware & power management & Sampling measurement & Intel and AMD CPU \\ 
        \hline
        ACPI\cite{acpi} & specification & power management & Dynamic Power Management, Power capping & x86 CPU \\ 
        \hline
        DT\cite{devicetree} & specification & power management & Dynamic Power Management, Power capping & ARM CPU \\ 
        \hline
        Perf tools\cite{perftools} & software & performance and energy management & query RAPL counters & Linux with Intel and AMD CPU for energy \\ 
        \hline
        PAPI\cite{PAPIenergy} & software & performance and energy management interface & query RAPL counters & All Linux systems \\ 
        \hline
        Likwid-powermeter\cite{likwid2010} & software & power profiling & query RAPL counters & Linux devices with Intel processor \\
        \hline
        PowerTOP\cite{powertop} & software & energy monitoring & query Intel RAPL & Linux with AMD or Intel devices \\
        \hline
        PyJoules\cite{pyjoules} & software & energy monitoring & query RAPL and Nvidia SMI interfaces & Linux with AMD, Nvidia or Intel devices \\
        \hline
        Powerstat\cite{powerstat} & software & measure energy consumption & query RAPL counters & Linux on Intel PCs \\
        \hline
        Power Gadget and PowerLog\cite{powergadget} & software & energy/power and temperature monitoring & query RAPL counters & Mac or Windows on Intel PCs \\
        \hline
        tx2mon\cite{tx2mon} & software & energy/power and temperature monitoring & query hardware counters & Marvell ThunderX2 \\
        \hline
    \end{tabularx}
   
    \label{tab:tab_CPU_tools}
\end{table*}
\end{center}

\subsection{Energy tools for microcontrollers}

- EEMBC CoreMark-Pro \cite{coremark} is a benchmark that aims at becoming the industry standard for embedded platforms. It contains five (resp. four) prevalent integer (resp. floating-point) workloads. The workloads in CoreMark-Pro represent a wide variety of performance characteristics, memory utilization, and instruction-level parallelism, highlighting the strengths or and weaknesses of the target processor in term of performance and energy efficiency.

- EEMBC ULPBench \cite{eembc2014} is a benchmark whose the goal is to overload a given processor in order to help determining the maximal amount of energy consumed. The benchmark consists of a number of mathematical and sorting operations. The STMicroelectronics PowerShield provides the backbone of the framework for probing an embedded system energy measurement.

- Dr. Wattson \cite{Drwattson} is an Energy Monitoring Module for high quality energy monitoring and measurements for microcontrollers boards. It is coupled with easy to use Arduino and Python libraries to provide quality AC energy data like {\em RMS Current}, {\em RMS Voltage}, {\em Power Factor}, {\em Line Frequency}, {\em Active/Apparent Power}, with just a few of lines of code.

- PSoC (Programmable System on Chip) 5LP\cite{Jankovi2015MicrocontrollerPC} is a data acquisition (DAQ) system for measuring and analyzing the power consumption of microcontrollers. DAQ system consists of a current measurement circuit using potentiostat technique (i.e, apply constant voltage during experiment). The DAQ device is based on system on chip PSoC 5LP and Python program for the analysis, storage and visualization of measured data. Implemented DAQ device is connected with a computer through a USB port and tested with developed Python program. 

- $N^3$ profiler \cite{DJEDIDI2020101805} is a power consumption monitoring tool to detect anomalies in power consumption for ARM-Based embedded systems at the level of the components. The authors used NARX (Nonlinear AutoRegressive eXogenous) \cite{en11030620} neural networks model as estimator to monitor energy/power for profiling and diagnosis purposes. $N^3$ improves upon the accuracy reported in the literature while maintaining low power and computational overhead. Experimentation was done on a smartphone considered as an embedded device.

\begin{center}
\begin{table*}
    \caption{SOTA Embedded systems energy/power management tools.}
    \begin{tabularx}{\textwidth}{|c|c|X|X|X|} 
        \hline
        {\bf Name} & {\bf type} & {\bf objective} & {\bf techniques} & {\bf portability} \\ 
        \hline
        EEMBC CoreMark\cite{eembc2014} & software & system benchmark for energy consumption & stress CPU with specific workload & 8 to 64-bit microcontrollers\\ 
        \hline
        EEMBC ULPMark\cite{eembc2014} & software & system benchmark for energy consumption & stress CPU with specific workload & 8, 16 and 32-bit microcontrollers\\ 
        \hline
        PSoC 5LP\cite{Jankovi2015MicrocontrollerPC} & software and hardware & system benchmark for energy consumption & SoC Module based on PSoC and python program & all microcontrollers \\ 
        \hline
        Dr. Wattson \cite{Drwattson} & software and hardware & Energy Monitoring & SoC Module based on Arduino and python program & Arduino, Raspberry and similar microcontrollers \\
        \hline
        $N^3$\cite{DJEDIDI2020101805} & software & Monitoring, diagnosis, software optimization & machine learning prediction & embedded systems \\
        \hline
    \end{tabularx}
    \label{tab:tab_Embedded_tools}
\end{table*}
\end{center}

{\bf Comment :} For a more accurate power measurement on micro-controller board, the following actions can be considered: disable HDMI and LEDs if present; minimize accessories usage (a connected keyboard for instance); be selective with Software (different programs running) and disable WiFi. Different system commands can be used depending on the micro-controller to disable the aforementioned features.

\subsection{Energy tools for Modern HPC systems}

- HDEEM (High Definition Energy Efficiency Monitoring)\cite{Hackenberg2014HDEEMHD} is an FPGA-based system on-chip that is intended to equip a compute node for its power measurement. The aim is to aggregate at high frequency (1 kHz) the measurements made by watt-metrics probes distributed among the components of the compute node. The samples associated to the last 7 hours of execution can be stored in a local memory of HDEEM for direct accesses through a programming interface in C language and/or through reads from report files.

- Similar to HDEEM, WattProf\cite{7307670} is a system-on-chip based on an FPGA that can be connected via a PCIe interface to a compute node. WattProf comes with dedicated wattmetric probes that can be plugged on the PCIe interface of the targeted hardware components and also on the connectors for the DRAM. WattProf includes a memory for storing samples of energy consumption measurements, and an API to access those samples.

- DiG (Dwarf in a Giant) \cite{Libri2018DwarfIA}, is another system on-chip based on an Arduino 5. Unlike HDEEM and WattProf, DiG connects to the power supply of the computer and thus captures its overall energy consumption rather than that of individual components. The Arduino board is used to process the energy consumption data, as well as to send them out through the network of the supercomputer. It might be more convenient or efficient to dedicate an individual unit to the management of the measurements coming from the participating compute nodes. That unit will thus serve as the provider of energy measurements to the user. In addition, DiG also allows for accurate and high frequency sampling, while remaining a low cost system-on-chip for HPC.

- PowerPack \cite{4906989} was the first tool to isolate the power consumption of common devices including {\em disks, memory, NICs}, and {\em CPU} within a given machine and correlate the corresponding measurements with the main subroutines of the applications being profiled. The framework support multi-core and multiprocessor-based nodes and provides in-depth analyses of the energy consumption of parallel applications. These analyses include the impacts of multiprocessing at the level of the chip on energy efficiency. The authors used the framework to study the power dynamics and energy efficiencies of DVFS techniques on clusters, and the experiments showed that DVFS scheduling can intelligently enhance system energy efficiency while maintaining processing  performance. They claim that their methodology as described in their work can be extended to other architectures and measurement devices. For instance, one can directly use the power sensors integrated in emergent computer systems for a more convenient power measurement.

- BDPO(Bull Dynamic Power Optimizer)\cite{DBLP:phd/hal/Stoffel21} is a dynamic reconfiguration tool that runs as a daemon behind a given HPC application and adapts the clock frequency of the CPUs according to the workload. It has the particularity of being completely agnostic to the considered application, as well as to the platform, while not requiring any configuration from the user. The authors of the tool experimentally got that the use of BDPO reduces the energy consumed by the execution of NEMO and HPCG applications by about 15\%, while maintaining the associated overhead below 4\% \cite{DBLP:phd/hal/Stoffel21}.

- Phase-TA \cite{DBLP:phd/hal/Stoffel21} is a tool for analysing the profiles of iterative HPC applications, especially those produced by Bull Dynamic Power Optimizer (BDPO) \cite{DBLP:phd/hal/Stoffel21} (see the previous paragraph). It detects locally periodic behaviours and try to characterise them by constructing patterns corresponding to the associated periodicities. The authors experimentally showed that the patterns constructed by Phase-TA are relevant representations of the considered periodicities, which seem to dominate the execution time. The observed performance of Phase-TA allows to consider the use of Phase-TA during the execution of an HPC application for its energy monitoring.

- GEOPM (Global Extensible Open Power Manager) \cite{Eastep2017GlobalEO} is provided as a runtime framework designed to address the power scaling challenge in Exascale systems. GEOPM's architecture is described as having a plugin extensible design, which enables a fast  prototyping of new energy management strategies. This flexibility allows different plugins to be tailored to the specific priorities of individual HPC centers. The package comes with a specific plugin for GEOPM, referred to as a "power rebalancing plugin.", which is designed for power-capped systems and aims at improving efficiency by minimizing job time-to-solution within a given power budget. Provided experimental results 
 demonstrate the potential of the GEOPM framework. The power rebalancing plugin reportedly achieves up to 30\% improvements in the time-to-solution of CORAL system procurement benchmarks \cite{Coral2015} on a Xeon Phi cluster. The authors present GEOPM as a contribution to the High-Performance Computing (HPC) community, and they encourage collaborations and accelerated advances in energy management solutions.

% - PMAC (Power Monitoring and Controlling) Tool \cite{pmac2022} is a web-based power monitoring and controlling tool for energy optimization of HPC applications. PMAC reports the power consumption of the software as well as for the hardware  in real-time. It allows to manages power based on application’s profile and DVFS mechanism. The specificity of the tool is that it can be used as an energy profiling as well as an energy optimizer. In the latter case, the tool uses its own profile report to guide the power optimization strategy. Experimental results have shown an energy saving of 12 -15\% when using P-MAC. P-MAC uses CMAF (C-DAC Multi-Agent Framework) for the transmission and execution of control policies.

% - EERT (Energy Aware Rescheduling Tool)\cite{eert2022} is another energy management tool that act on the internal scheduling of HPC applications in order to the reduce energy consumption through maximizing the CPU utilization and  switching off idle nodes. The benefit is more noticeable when  there is a important imbalance in the distribution of workloads over the nodes. EERT seamlessly uses the {\em distributed multithreaded check-pointing (DMTCP)} mechanism for check-pointing. Experimental results provided by the authors show ~15\% energy saving when using EERT.

- EAR (Energy Aware Runtime) \cite{lenovo2022} is an energy management framework for {\em energy measurement, energy management} and {\em energy optimization}.  EAR supports standard CPUs as well as (NVIDIA) GPUs. It is constantly being enhanced to support other and upcoming technologies as well. The optimization of the energy consumption of an HCP cluster is done at two levels: the {\em compute node level}, which is provided by the EAR library and the {\em system level} for power capping using DVFS techniques.

- FIRESTARTER \cite{9555945}, is a handy utility that aims at creating near-peak power consumption on standard compute nodes. It can be used for tests of cooling and power infrastructures, system stability test, or as a maximum power consumption baseline for application energy efficiency studies. FIRESTARTER is currently only available for the Linux operating system and has supports for Intel architectures (Nehalem, Westmere, Sandy Bridge, Ivy Bridge, Haswell, Broadwell, Skylake, Knights Landing), AMD family 15h and 17h (Zen, Zen+, Zen 2) processors, and NVIDIA GPUs. The tools stresses the most important {\em power consumer} parts of compute nodes: CPU (cores + memory related components such as the caches), GPUs, and main memory and report some metrics that include power consumption.

- lo2s \cite{lo2s} is a lightweight performance monitoring tool. The tool collects performance and energy data w.r.t  various metric (i.e., {\em perf counters, kernel trace-points, model specific registers}, and {\em custom metric data} provided by related plugins). These trace data are stored in the {\em Open Trace Format 2} (OTF2), that can be used for offline analysis with tools like Vampir \cite{vampir}. Ilsche et al.\cite{lo2s2018} investigated improvements of lo2s by combining a detailed recording of system events with information from a high-resolution power measurement while recording the applications execution trace and C-state transitions.

- READEX \cite{Oleynik2015RunTimeEO} is a support tool suite to improve the energy-efficiency of HPC applications. It allows to dynamically adjust energy-related characteristics at runtime according to the actual resource requirements and thereby improves energy-efficiency and performance. It uses a multi-agent based approach to identify runtime situations and to determine optimal system configurations. The tools also provides insights for the specification of domain knowledge to improve the automatic tuning impact. The result of the analysis step guides runtime tuning. 
%%%%%%%%%%% COMMENTED ON THE WAY TO IEEE Submission%%%%%%%%%%%%%%%%%%%%%%%%%%%
%Figure \ref{fig:READEX} provides an overview of READEX working diagram.
%\begin{figure}[ht]
%    \centering
%    \includegraphics[scale=0.3]{images/READEX-diagram.png}
%    \caption{READEX working diagram}
%    \label{fig:READEX}
%\end{figure}
%%%%%%%%%%%%%%%%%%%%%%%%%%%%%%%%%%%%%%

- MERIC \cite{Vysocky2017MERICAR, Schuchart2017TheRF} is Lightweight C/C++ library (with an interface for Fortran applications) that measures energy consumption and timings of annotated regions inside a user application. The MERIC library evaluates application behavior in terms of resource consumption and runtime parameters including {\em Dynamic
Voltage and Frequency Scaling} (DVFS) and {\em Uncore Frequency Scaling} (UFS). It performs dynamic application tuning following the READEX approach. The library was originally developed for Intel x86 systems, but additional supports for AMD, IBM and selected ARM systems chips was added. It also supports HDEEM and DiG hardware tools for energy measurement.  A tool called RADAR VISUALIZER \cite{Vysocky2017MERICAR} for visualization of the analyzed application behavior in different system configurations was proposed to analyze MERIC results.

\begin{center}
\begin{table*}
    \caption{SOTA Supercomputers systems tools for energy/power management}
    \begin{tabularx}{\textwidth}{|l|l|X|X|X|} 
        \hline
        {\bf Name} & {\bf type} & {\bf objective} & {\bf techniques} & {\bf portability} \\ 
        \hline
        WattProf\cite{7307670} & hardware & energy/power measurement & system on-chip based power monitoring board & all server node \\ 
        \hline
        HDEEM\cite{Hackenberg2014HDEEMHD} & hardware & power measurement & system on-chip based power monitoring board & all server node \\ 
        \hline
        DiG\cite{Libri2018DwarfIA} & hardware & energy monitoring & system on-chip based power monitoring board & all server node \\ 
        \hline
        PowerPack\cite{4906989} & software & energy/power measurement & isolate power consumption of devices in measurement & Linux systems \\ 
        \hline
        GEOPM \cite{Eastep2017GlobalEO} & software & power management & dynamic rescheduling, power rebalancing & Linux HPC systems\\ 
        \hline
        Phase-TA\cite{DBLP:phd/hal/Stoffel21} & software & energy profiling & analysing the profiles of HPC applications & Linux systems \\ 
        % \hline
        % PMAC\cite{pmac2022} & software & power management & DPM and DVFS techniques; web based monitoring & Linux systems \\ 
        \hline
        BDPO\cite{DBLP:phd/hal/Stoffel21} & software & power optimization & DFS on computing cores during workloads execution & Linux x86 systems\\ 
        \hline
        lo2s\cite{lo2s} & software & performance and energy profiling & Sample hardware counters events & Linux x86 systems \\ 
        \hline
        EAR\cite{lenovo2022} & software & energy management & DPM techniques, power capping, On/Off policies & Linux with Intel, AMD and Nvidia devices \\ 
        \hline
        READEX\cite{Oleynik2015RunTimeEO} & software & energy and performance optimization & exploit the dynamic behaviour of application and make resources allocation & Linux x86 and ARM systems \\ 
        \hline
        MERIC\cite{Vysocky2017MERICAR} & software & energy management & dynamic application tuning and hardware energy measurement & Linux x86, ARM and Nvidia GPUs systems; HDEEM and DiG supports \\ 
        \hline
        FIRESTARTER\cite{9555945} & software & benchmark tests of cooling and maximum power consumption & stress execution units and data transfer between cores and memory hierarchy & x86 CPU and GPU \\ 
        \hline
    \end{tabularx}
    \label{tab:tab_HPC_tools}
\end{table*}
\end{center}

\section{About cooling Systems}\label{sec:cooling}
Designing computers that perform tasks efficiently without overheating is a major consideration for all manufacturers nowadays. Current CPUs and GPUs has a power consumption from tens to hundreds watts. Some specific CPUs consume little power like those of embedded systems and mobile devices (few milliwatts or microwatts).
Computers consume electrical energy and dissipate part of it in as heat coming from the resistance in the circuits. Excessive heat is a clear threat for the integrity of hardware components, with the risk of leading to serious damage. Thus, {\em cooling system}, which can be internal or external, is crucial in order to cap cap the dissipated heat so as to avoid a critical overheating.

\subsection{Cooling technologies for HPC systems} 
Cooling is crucial to HPC systems, especially for large-scale ones, but choosing the right technology depends on several factors like the {\em desired temperature} limits and the {\em operating cost}. There are mainly four types of cooling that are commonly considered: 
\begin{itemize}
    \item {\bf{\em air cooling}}: is the most basic cooling mechanism and also the most used one.
    \item {\bf{\em liquid cooling}}: needs less energy to operate and more efficient \cite{liquid_cooling}.
    \item {\bf{\em rear door heat exchanger (RDHX)}}: combines both air and water cooling mechanisms, with a great efficiency on data centers \cite{Schmidt2009ServerRR}. Its strengths include {\em Energy Efficiency},   {\em Cooling Efficiency},  {\em less space}, and  {\em flexibility}  \cite{RDHX-benef}.
    \item {\bf{\em immersive cooling}}: considers a direct immersion of the hardware in a dielectric (but thermally conductive) liquid (also called {\em coolant}) \cite{PAMBUDI20229509}. While it has the potential to deliver the highest performance and PUE, it can make the replacement of components tricky\cite{cooling}.
\end{itemize}

\subsection{News trends in liquid cooling design} 

The market for high-performance cooling systems has grown significantly as technology has shifted from simple air cooling to solution using liquid (including immersion cooling). Water is still the standard for most HPC users as it provides a good balance between performance and set-up cost.

While there are various variants, the basic concept remains the same. Water is pumped through a closed system up to a back plate placed near the hot components. Water has a better thermal conductivity, so this is potentially a higher performance system, but requires additional infrastructure. For example, many water-cooled data centres have a raised floor, so so that all the pumps can be conveniently routed and driven to the targets.

Meyer and Wettig. \cite{Meyer2013iDataCoolHW} developed {\em iDataCool}, an HPC architecture based on {\em IBM’s iDataPlex} platform, whose air-cooling solution was replaced by a custom water-cooling solution. A significant portion of the energy spent on HPC systems can be recovered in the form of chilled water, which can then be used to cool other parts of the computing center. The authors illustrated the cooling performance and the energy reuse efficiency through benchmarks.

Nonaka et al. \cite{Nonaka} provided a quantitative and systematic analysis of the impact of the cooling water temperature on HPC infrastructures. They evaluated the efficiency of the hot water cooling technique, taking into consideration not only the energy reduction on the facility side (cooling system), but also the impact on the power consumption and on the performance degradation from the machine side. They showed that, contrary to the gain in the energy consumption, on the HPC facility side, when using higher temperature cooling water, there is an increase in the number of nodes suffering from performance degradation, especially at synchronization barriers.

Ljungdahl et al. \cite{LJUNGDAHL2022117671} developed a {\em decision support model} that takes basic information of a given cluster or data center as inputs and provides a parameterized output that shows various configurations and design parameters that can be considered. The main outputs include {\em energy savings, cost savings} and {\em efficiency gains} through the Power Usage Efficiency(PUE) and the Energy Reuse Efficiency(ERE). An electricity saving between 8.14\% and 10.8\%  and a waste heat recovery of 85 to 576 MWh/year were obtained in a Danish case study. Additional system configurations beside  existing local heating source showed an energy saving of 32\%. The goal of the decision support model is to assist the design of future waste heat recovery applications through selection of system parameters including coolant temperatures, energy storage design parameters, District Heating supply temperatures and District Heating load coverage from the data center or HPC cluster.

\section{Energy Optimization techniques}\label{sec:optimization}

\begin{figure*}[h]
  \centering
  \includegraphics[scale=0.25]{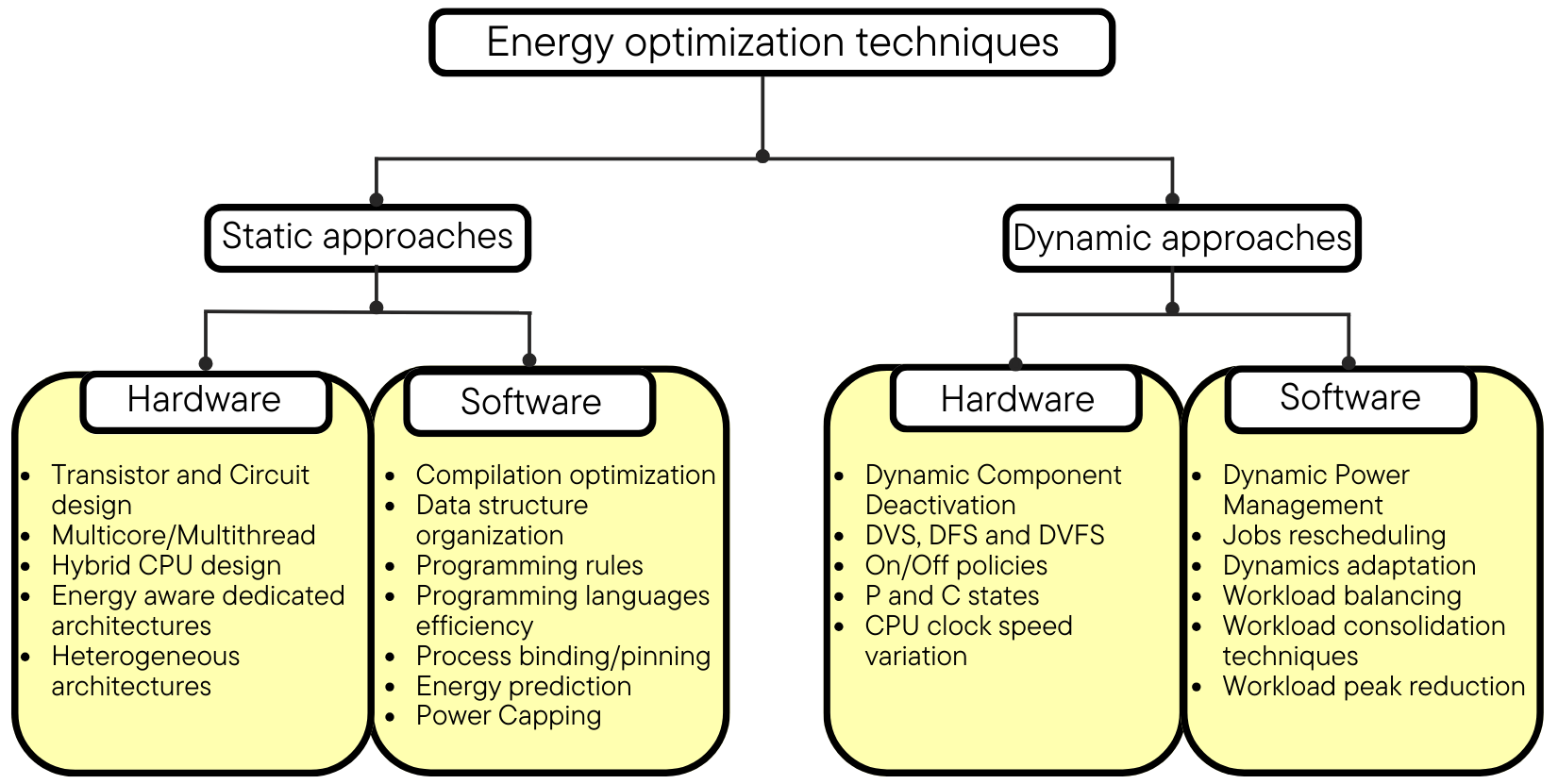}
  \caption{Taxonomy of power/energy optimization techniques in computer system.}
  \label{fig:optimize}
\end{figure*}

The power optimization techniques aim at minimizing the energy consumption besides traditional metrics like computing time or memory space. This concern is crucial when there is a power constraint as when the available energy is limited or its supply is costly. Power optimization can be addressed through hardware and software approaches, considering static or dynamic strategies. Figure \ref{fig:optimize} gives an overview of existing energy optimization techniques grouped by their nature.

\subsection{Static energy optimization approaches} %State-of-the-art

{\it - Hybrid CPU design} : Hybrid design in CPU is an approach that combine low power/performance and high power/performance cores. This was introduced by ARM with BIG.LITTLE architecture and similarly considered more recently by Intel with "Lakefield" chip \cite{Lakefield}. Intel 12th generation CPUs (family code name: "Alder Lake") are designed following this hybrid model for energy saving and battery long life for laptop computers.

{\it - Programming languages efficiency} : Pereira et al \cite{10.1145/3136014.3136031} studied the energy efficiency of 27 programming languages, monitoring their performance using ten different programming problems. Out of these 27 selected languages, Python ranked 26. Python used 59x more energy than the most efficient language, which is the C language. Nowadays, Python might be the best choice is many cases, for instance when building and training neural networks. There is a significant potential energy saving when considering a more energy efficient language. The authors showed interesting findings such as slower/faster languages consuming less/more energy and how memory usage influences energy consumption.

{\it - Programming aspects} : A practical approach to C++ was presented in the work by Meyers et al. \cite{meyers2005effective} that describes the basic rules followed by experts (i.e., the things they do or avoid as much as possible) to produce clear, correct, efficient code. This is a static optimization technique for time-to-solution and energy-to-solution by considering the best programming practices.

{\it - Machine learning prediction models} : Gao et al. \cite{Gao2014MachineLA} developed a neural network framework that learns from actual operating data to model plant performance and accurately predict the PUE. The results demonstrated that machine learning is an effective way of leveraging existing sensor data to model DC performance and improve energy efficiency.

{\it - Circuit design} : the work by Beloglazov et al. \cite{article_2} presents Static Power Management (SPM) techniques that contains all optimization methods applied at the design time at various levels including {\em circuit, logic, architecture} and {\em system}. Circuit level optimizations consist in the reduction of switching in logic-gates and combination circuits through a complex gate design and transistor sizing.  

{\it - Energy-aware dedicated architectures}: This category is for hardware level methods, which consider incorporating power optimization in the design process \cite{article_2}. In other words, an efficient mapping of high-level specifications into the design of the chip is applied. Apart energy-aware hardware design, it is  important to carefully consider a skillful programming that efficiently take into account the energy specificities of the target system. Most often, dedicated architectures (GPU, FPGA, TPU, etc.) are used for specific kernels based on the aforementioned observations.

{\it - Analytical models prediction for scheduling} : Tadonki et al. \cite{Claude2003} designed a combinatorial energy-ware methodology for an efficient management of power states of the RAM. The authors considered traditional techniques like tiling to improve the efﬁciency of the proposed methodology. Experimental results through simulations considering two well-known algorithms ({\em Transitive Closure} in graphs theory and {\em Fast Fourier Transform}) illustrated the efficiency of their strategy,  with 98\% energy reduction related to memory accesses for the Transitive Closure \cite{tado_europar}. Another analytical model for memory energy minimization was proposed by Tadonki and Rolim  \cite{claude2004} based on a formal model that captures the relation between energy and memory mechanisms into a {\em mathematical programming} form. They successfully evaluated their approach considering the model of a standard RDRAM so as to figure out the behavior of each parameter together with the energy that can be saved or lost. Singh et al. \cite{Mitali2003} developed algorithmic techniques for memory energy reduction by exploring the structure and data access pattern to devise an efficient memory power management schedule. They investigated and discussed optimality considering theoretical lower bounds on memory energy. Simulations results demonstrated significant energy reduction over other existing approaches.

{\it - Compilation and programming best practices} : Quantitative modeling of energy prediction is one of the more attractive approach for static performance optimisation, as it could be used for power aware programming or compilation. Power optimization schemes can be incorporated into compilers by exploiting recurrent patterns in programs\cite{Lin2013} and the energy cost of individual programming instructions. For the latter, Leite et al. \cite{LEITE20142260} proposed a fine grained approach for power analysis and prediction, with a focus on a set of basic programming instructions ({\it addition, multiplication, division, memory read, memory write, memory copy, print, comparison, malloc}). The authors used a series of micro-benchmarks to measure the energy cost per operation considering both the overhead of the embedding loop and that of associated optimizations. Their results showed a 9.48\% error rate on the energy prediction of a sorting algorithm. However, their work do not consider specific instructions like the FMA (Fused Multiply Add), which is now available in most of the major CPUs for floating-point operations performance consideration.

{\it - Power Capping} : Power capping is a technique used for setting a power threshold to not be exceeded by the considered hardware unit. Manufacturers use to provide appropriate tools to handle power capping: NVIDIA System Management Interface (NVIDIA), RAPL (Intel), Energyscale (IBM), and APM (AMD). Cabrera et al. \cite{inproceedings_2} performed an analysis of the performance and the energy efficiency using power cap technologies with a selection of various applications. They illustrated the case of the Intel power cap and the NVIDIA power limit technologies. They extracted a Pareto front of the configurations that lead to the most efficient energy usage for the best possible performance. They provide a methodology base on energy-performance trade-off selection. But there is no investigation on how to automatically select the best trade-off.

\subsection{Dynamic energy optimization approaches} %State-of-the-art
{\it - DPM (Dynamic Power Management) techniques} : DPM techniques are approaches that include methods and strategies for run-time adaptation of a system's behavior according to current resource requirements or any other dynamic characteristic of the system's state. A paper by Beloglazov et al. \cite{article_2} presents these techniques with the assumption that enabling DPM allows dynamic adjustment of power states according to current performance requirements. Another assumption is that the workload can be predicted. The authors described different levels for DPM techniques based on hardware (resp. software) considerations. Hardware DPM can be different from one hardware component to another, but they are usually classified as Dynamic Performance Scaling (DPS) such as {\em Dynamic Voltage and Frequency Scaling (DVFS)} and partial or complete {\em Dynamic Component Deactivation (DCD)} during idle periods. Some software DPM techniques (Intel RAPL and Nvidia-msi for instance) apply hardware DPM in accordance with the system's power management.

{\it - DFS (Dynamic Frequency Scaling) and DVS (Dynamic Voltage Scaling)} adjust the frequency and the power of computing devices (i.e, CPU, GPU, FPGA) by scaling the clock frequency/voltage  according to the execution of memory or compute-bound application kernels \cite{Sueur2010DynamicVA}. So a significant reduction of the total power consumption can be achieved with different voltage/frequency reduction levels. But very often, voltage and frequency ranges are fully interdependent, (i.e., a change in clock frequency does imply changes in the supply voltage, and vice versa). For this case, DVFS was proposed.

{\it - DVFS (Dynamic voltage and frequency scaling)} : DVFS is a technique to systematically adjust the power through a dynamic adjustment of both voltage and frequency settings of a computing device, controller chips and peripherals in order to optimize resources allocation for the tasks and maximize power saving when those resources are not needed. Morán et al. \cite{inproceedings_1} evaluated a series of strategies that can be applied to improve energy efficiency when a failure occurs. This strategy uses the {\em Advanced Configuration and Power Interface (ACPI)}. They considered the use of DVFS techniques and system hibernation at the node level. They estimated the execution time and the waiting time of processes that do not fail through a characterization of the energy consumption required to execute the application and its communication pattern. They considered a simulator to conduct their experiments so as to not get bothered by the issues of a real system and thus focus on the essential.

{\it - Workload consolidation techniques} : Sanjeevi et al.\cite{Sanjeevi2017} proposed an extensive  background and motivation of workload consolidation techniques in the cloud computing context \cite{Sanjeevi2017, leite2014excalibur}. In their paper, they described four recent workload consolidation algorithms considering the goal of  reducing energy. In addition, the features of the best workload consolidation algorithm is highlighted.

{\it - On/off policies} : shutdown policies stand an appealing approach to dynamically adapt the active resource configuration to the actual workload by tuning off unused components in order to reduce power consumption. However, there are some important constraints to take into account for these policies like the cost (time and energy) of switching on and off, the power and energy consumption bounds caused by the electricity grid or the cooling system, and the availability of renewable energy. Benoit et al. \cite{benoit:hal-01557025} studied the existing approaches that are based on these policies and proposed models for translating the energy constraints into different shutdown policies that can be combined for a multi-constraint purpose. 

{\it - Workload peak reduction} : Sai et al.\cite{6560768} presented a  space-time multiplexing (STM) power management technique implemented through DVFS for workload balancing. The physical design parameters are based on 130nm CMOS process with TSV models. Experiment results showed that their  approach can lead to a peak of 38.10\% power reduction and 2.60x workload balancing.

\subsection{Hybrid energy optimization approaches}

Vaddina et al.\cite{DBLP:conf/iscc/VaddinaLO21} proposed a workflow for energy and temperature profiling on systems running parallel applications. They did their experimentations standard multi-core processors using common benchmark applications. Their strategy allows full and dynamic runtime control so as to keep the frequency of the processors within a predetermined range. By this way, they showed that the energy response to frequency scaling is highly dependent on the workload characteristics and it is a convex fonction around the optimal frequency point. Another interesting result from their work is the fact that the tested low-power processor was consuming more power on average than the other standard processors. Their investigation surely contribues to the understanding of power dissipation and its link with temperature as the necessary first step towards optimizing the energy efficiency of HPC systems.

Grant et al.\cite{8323587} presented a taxonomy of power profiling techniques on modern HPC platforms. The authors used three HPC mini-applications for analysis on three production HPC systems to examine meaningful details, scope, and complexity of the selected energy profiling techniques. Their work demonstrates that a combination of out-of-band measurement with in-band profilers can provide a detailed and accurate view of power usage with almost no overhead.

Jafari-Nodoushan et al. \cite{9118980} proposed a heuristic battery-aware scheduling policy for periodic and non-periodic real-time tasks under DVS mechanism, with an explicit consideration of power leakage. They compared the battery consumption of their proposed policies with an optimal solution, which could be derived via Calculus of Variations (CoV). Experimental  results showed a maximum of 17.7\% (resp. 11.3\%) battery charge saving for non-periodic (resp. periodic) tasks in comparison to the critical frequency method.

\section{Energy of AI processing}\label{sec:ai_concern}
\subsection{Motivation}
Training a single AI model can emit as much carbon as five cars in their lifetimes \cite{Wu2021SustainableAE}. Yet, this analysis pertained to only a one-time training run. When the model is improved by repetitive training, the energy cost is significantly greater. Many large companies, which can daily train thousands and thousands of models, are taking the energy issue more seriously. The work by Strubell et al.\cite{inproceedings_3} describes and analyses the problem by exploring AI’s environmental impact, studying ways to address it, and issuing calls to action.

Cutting-edge AI models have nowadays billions of parameters and more. One popular case, GPT-3, has 175 billions of machine learning parameters. The model was trained on NVIDIA V100, but researchers have estimated that the full training would have cost 34 days and \$4.6 millions with 1024  A100 GPUs. While energy usage has not been disclosed, it’s estimated that GPT-3 consumed 936 MWh\cite{torwald2021}. As the models get bigger and bigger in order to handle more complex tasks and the huge volume of requests, the demand for high-end servers to process the models grows exponentially.

Since 2012, the computational resources needed to train cutting-edge AI systems have been doubling every 3.4 months \cite{openAI2018}. This escalation in energy use/requirement stands against the common promise of reaching carbon neutrality in the coming decade.

\subsection{Studies on CO$_2$ concerns with AI}

Qiu et al.\cite{qiu2022look} provided a pioneer systematic study of the carbon footprint of federated learning. They proposed a rigorous model to quantify the carbon footprint, thereby facilitating any investigation of the relationship between federated learning design and carbon emissions. They showed that federated learning can emit up to two orders of magnitude more carbon than centralized machine learning. However, in some settings, both approaches can be comparable because of the low energy consumption of embedded devices. Their work highlighted future challenges and trends in federated learning about  reducing its environmental impact considering algorithms efficiency, hardware capabilities, and stronger industry transparency.

Luccioni et al.\cite{Luccioni2022EstimatingTC} provided an estimate of the carbon footprint of BLOOM, a 176-billion parameter language model, over its lifetime. The authors estimated that BLOOM’s ﬁnal training emitted approximately 24.7 tons of $CO_{2}e$ for the dynamic power consumption only, and 50.5 tons for all processes ranging from equipment manufacturing to the operational phase. The energy requirement and carbon emission of its deployment for inference via an API endpoint receiving user queries in real-time was also studied. The authors also discussed the difﬁculty of estimating accurately the carbon footprint of ML models and reported future research directions that can contribute to  improving carbon emission.

Patterson et al. \cite{Patterson2021CarbonEA} studied the carbon footprint of large-scale neural network training and discussed about opportunities to improve energy efficiency and $CO_{2}$ emission. The authors estimated the energy consumption and the carbon footprint of several recent large models: T5, Meena, GShard, Switch Transformer, GPT-3, and Evolved Transformer. Their study illustrate that the choice of {\em neural network architecture, datacenter}, and {\em processing unit} can reduce the carbon footprint by ~100-1000x. The authors also highlighted the need for more focus on how to improve emission metrics in addition to accuracy. Addressing these concerns lead to a reduction of the carbon footprint of ML through accelerating innovations in the efficiency of the algorithms, systems, hardware, datacenters, and in carbon free energy.

Wu et al.\cite{Wu2021SustainableAE} studied optimizations techniques for operational energy footprint reduction across Facebook’s AI applications. Their work showed improvements on different standpoints: model, platform, infrastructure, and hardware. They described optimization techniques on Platform-Level Caching and showed an improvement of power efficiency by 6.7× with application-level caching. Another optimization is GPU acceleration. In addition to caching, deploying across GPU-based specialized AI hardware unlocks an additional 10.1× energy efficiency improvement. Algorithmic optimizations provided an additional 12× energy efficiency reduction. Considering half-precision (e.g., going from 32-bit to 16-bit operations) provided a 2.4× energy efficiency improvement on GPUs. Another 5× energy efficiency gain was achieved by using custom operators to schedule encoding steps within a single kernel of the Transformer module.

Patterson et al.\cite{Patterson2022} presented some best practices to reduce the energy of ML training by up to 100x and $CO_2$ emissions by up to 1000x. The authors recommended that ML papers should explicitly include $CO_2e$  to foster competition not only on model quality. Publishing emissions ensures accurate accounting. They showed that for large-scale ML deployments, minimizing emissions from training should not be the unique as subsequent steps like serving also count. Approaches like {\em neural architecture search} increases emissions but lead to more efficient serving and a strong overall reduction of the carbon footprint of ML. The work also highlighted the carbon footprint to be erased entirely if cloud providers could fully consider renewable energy (it is already the case with Google and Facebook, and will soon be the case with Microsoft Azure). Another interesting insight from their work is that published studies overestimate the cost and carbon footprint of ML training because they didn't have access to exact information or because they extrapolated point-in-time data without accounting for algorithmic or hardware improvements.

Ludvigsen \cite{Ludvigsen2022} demonstrated the difficulty in determining the environmental impact of Machine Learning as a field. Moreover, he showed how easy it is for practitioners to estimate the carbon footprint of their machine learning models with tools like {\em CodeCarbon}\cite{codecarbon} or {\em ML CO2 Impact}\cite{lacoste2019quantifying}. In addition, 17 concretes ideas on how to reduce the carbon footprint of machine learning models are also presented. Some of these ideas can be easily implemented, while others require more efforts and expertise. Indeed the energy profiling of AI applications is a serious focus worth investigating \cite{youssef2,youssef4}.

The ecosystem of sustainable AI is presented and commented by Zhao et al. \cite{youssef1}. They presented an overview of various areas for potential changes and improvements from the standpoint of operational and hardware optimizations for HPC systems considering AI workloads. Three aspects covering the main issues from a micro-to-macro perspective analysis are proposed: infrastructure and resource utilization, user and behavior, and the community of the researchers and practitioners. They showed that concerted and unified efforts are required in order to make effective the transition to a greener ecosystem for AI researches and practices.

\subsection{Energy profiling tools for AI applications}
Several tools have been developed in recent works about estimating the carbon footprint of machine learning models. These tools estimate the carbon footprint from energy consumption {\em estimates} or {\em measurements}.

\subsubsection{Tools that operate from energy estimates}\
\\
{\em - ML CO2 Impact} \cite{lacoste2019quantifying}: This is a tool that calculates the amount of raw carbon emissions and an estimate of the offset carbon emissions. The latter value depends on the grid used by the cloud provider. About the estimation, it does not take into account the datacenter PUE (Power Usage Effectiveness).
\\
{\em - Green Algorithms} \cite{Lannelongue2020GreenAQ}: An online tool which enables users to estimate and report the carbon footprint of their computation. The tool easily integrates with given computational processes as it requires minimal information  and does not interfere with the considered code, while also accounting for a broad range of hardware configurations. With  power-hungry and expensive training algorithms coming from cutting-edge AI, the tool is worth considering for the address the underlying energy concern. 

\subsubsection{Tools that operate from energy measurements}\
\\
{\em - Codecarbon} \cite{codecarbon}: {\em Codecarbon} is a (lightweight) Python package that estimates estimates the amount of carbon dioxide (CO2) produced by a given code. It achieves that purpose by estimating the electricity power consumption (GPU + CPU + RAM) of the device and weighting it with the local carbon intensity (i.e. where the computing is actually done). The tool thus enables developers to track CO2 emissions across ML experiments or other programs. Power consumption will be successfully tracked only if there are RAPL files within the indicated directory. If not found, CodeCarbon will switch to a {\em fall back} mode. 

{\em - Tracarbon} \cite{Tracarbon}: {\em Tracarbon} is a Python library that tracks the energy consumption and thereby estimates carbon emissions. It detects automatically the key information like the location and the hardware type before starting the tracking. Tracarbon is a flexible tool designed to easily include other platforms, cloud providers, carbon emission APIs, or other data exporters through a Command-line interface (CLI) with already defined metrics or programmatically with the API by defining the desired metrics. 

{\em - Eco2AI} \cite{budennyy2023eco2ai}: {\em Eco2AI} is a python library for $CO_2$ emission tracking. It monitors energy consumption of CPU and GPU devices and estimates equivalent carbon emissions by taking into account the local carbon intensity.  Eco2AI is applicable to any python script. All emissions data together with information about the device are recorded in a local file.

{\em - Experiment-impact-tracker} \cite{henderson2020systematic}: {\em Experiment-impact-tracker} is defined as a toolkit for tracking energy, carbon, and compute metrics for machine learning (or any other) experiments. The tool runs under Linux system on Intel chips and NVIDIA GPUs for which it records information related to carbon emissions.

{\em - Carbontracker} \cite{anthony2020carbontracker}: {\em Carbontracker} is a tool for tracking and predicting the energy and carbon footprint associated to the training of deep learning models. The output result includes {\em duration, energy}, and {\em carbon footprint} of training a given deep learning model with the main parameter (specified by the user) being the {\em number of monitored epochs}. The tool forecasts the {\em carbon intensity} related to the electricity production during the predicted duration, then uses it to predict the carbon footprint. At the preliminary stage of the development of the tool, a basic linear prediction model is considered.

{\em - Zeus} \cite{zeus-nsdi23}: {\em Zeus} is an online optimization framework for DNNs (Deep Neural Network) training workloads. The tool provides the Pareto frontier for energy-time consumption trade-off and allows users to navigate around by automatically tuning the batch size and GPU power limit of their jobs. Zeus uses an online exploration approach in conjunction with just-in-time energy profiling, thus overcoming the need for offline measurements, while adapting to data drifts over time. The authors shows that Zeus can improve the energy efficiency of DNNs training by 15.3\%–75.8\% for diverse workloads.

\subsection{Energy optimization of AI applications}

Cutting-edge AI models require a huge number of parameters and imply a noteworthy computing load, making them being considered as cumbersome from the standpoint of the classical complexity references (running time, memory, and energy). In addition, AI applications are expected to be intensively used both at the level of a single user for routine issues or a collective level (i.e. server mode). Thus, there is clear need for optimization techniques for the implementation (training and inference) and the deployment of large AI models on low-power devices considering their limited hardware characteristics. This section enumerate and describe the  mains techniques considered in the literature to cope with energy issues related to AI applications (design, implementation and execution). Those techniques can be grouped considering the following major categories: {\em Quantization, pruning, filters compression, matrix factorization, neural architecture search, knowledge distillation}, and {\em hardware selection}. 

\subsubsection{Quantization} \
\\
One of the most popular energy-aware approach of deep learning optimisation is {\em quantization}. Quantization is a technique to reduce the computational and memory costs by considering low-precision data types (e.g. 8-bit integer) instead of the ordinary ones (e.g. 32/64-bit floating point). For instance, inference could be implemented by representing the weights and activations with low-precision data types. Thus, quantization stands as a technique to speed up inference through running with quantized operators. Quantization can be also considered for the training phase in a so-called {\em quantization-aware training} approach. 
There several advantages of quantization including: 
{\em more compact model representation; wider vectorization (SIMD);  less memory storage; less energy consumption (potentially), faster computation, deployment on embedded devices}. Popular Deep learning frameworks like TensorFlow and PyTorch provide a quantization API to simplify the quantization process. Gholami et al. \cite{youssef3} provided a survey of quantization techniques for efficient deep neural networks.

\subsubsection{Pruning} \
\\
Pruning is technique applied on inference to get models of smaller in size, thus, similarly to quantization, it yields to better memory/energy/processing complexity  with minimal loss in accuracy. Removing less important parameters and connections from an original deep neural network can clearly reduce the volume of memory accesses and associated computations. In addition to the aforementioned advantages, pruning might allow for the execution of the considered model in low-end devices such as mobile/embedded devices. Yang et al. \cite{Yang2016DesigningEC} have shown in their work that an energy-aware pruning technique for AlexNet and GoogleNet can reduce energy consumption by 3.7.

\subsubsection{Filters compression} \
\\
Convolution kernels are the bulk of the computations in DNNs, and the fully connected layers contain around 89\% of the parameters (e.g.  AlexNet\cite{Goel2020ASO}). To reduce the power consumption of DNNs, the main focus has been on reducing {\em the number of arithmetic operations} (resp. {\em parameters}) in the convolution layers (resp.connected layers). The so-called {\em bottleneck architecture} \cite{bottleneck_arch} can significantly reduce memory and latency requirements of DNNs. Filter compression is orthogonal to pruning and quantization techniques. The three aforementioned techniques can be used together for a combined optimization approach to reduce energy consumption.

\subsubsection{Neural architecture search} \
\\
There are many different network architectures and optimization techniques to consider when designing low-power AI applications. Neural architecture search (NAS) is a technique for automating the design of artificial neural networks (ANN).
%It is often difficult to manually find the best AI for a particular task when there are many architectural possibilities. These aim to find in an automated way the right architecture, under given constraints of model size, depth, and/or width. Neural Architecture Search (NAS) and Automated Machine Learning are techniques that automates AI architecture design for various tasks. NAS uses a Recurrent Neural Network (RNN) controller and uses reinforced learning to compose candidate AI architectures. These candidate architectures are trained and then tested with the validation set. The validation accuracy is used as a reward function to then optimize the controller’s next candidate architecture.
Many works address the reduction of computational cost and environmental impact of deep learning by accelerating neural network architecture search and hyperparameter optimization. Frey et al. \cite{Frey2022} introduced a framework called {\em training performance estimation} (TPE), which is based on existing techniques for estimating the speed of the training process. Ranking models (by estimated performance) without training to convergence leads to a potential saving of up to 90\% of time and energy of the full training budget. Some variants of {\em early stopping} that generalize common regularization technique to account for energy costs were also proposed, and this approach enables significant energy savings across the entire pipeline of model development and deployment.
Narsin et al. \cite{Nasrin2022} proposed ENOS (Energy-Aware Network Operator Search in DNNs) approach to address the {\em energy-accuracy trade-off} of a deep neural network acceleration. The search in ENOS is formulated as a {\em continuous optimization problem} that is solvable using gradient descent methods. This lead to a minimal overhead in the training cost when learning both layer-wise inference operators and weights. ENOS improves accuracy by 10–20\% in comparison to the conventional uni-operator search approaches {\em under the same energy budget}. ENOS also outperforms the accuracy of comparable mixed-precision uni-operator implementations by 3-5\% for the same energy budget.
Some other works  based on the {\em splitting steepest descent} algorithm for fast energy-aware neural architecture optimization were also proposed \cite{wang2019energy, wu2019splitting}. 
 
\subsubsection{Knowledge distillation}\
\\
Knowledge distillation refers to the approach of transferring the knowledge from a large but unwieldy model or set of models to a single smaller model that can be deployed under real-world constraints.
There are three different ways that the larger teacher model ({\em teacher/student} analogy) is used to help training the smaller student model: {\em response-based knowledge, feature-based knowledge} and {\em relation-based knowledge} \cite{distil2021}.
Through a varying combination of these three techniques, it has been shown that some very large models can be migrated to smaller representations. Probably the most well-known of these is DistilBERT \cite{Sanh2019DistilBERTAD}, which is able to keep 97\% of its language understanding versus BERT, while having a model which is 40\% smaller and 60\% faster.

\subsubsection{Hardware selection for training} \
\\
Using processors that are optimized for ML training such as tensor processing units (TPUs) and recent GPUs (e.g. V100 and A100) instead of general-purpose processors can improve performance/watt by factors from 2 to 5 \cite{Patterson2022}.

\section{Conclusion}\label{sec:conclusion}
Energy concerns have an increasing priority for mainly two reasons. The first reason comes from the standpoint of “energy as a cost and/or constraint”. The cost of the necessary energy to keep HPC systems running with all surrounding aspects including cooling is becoming significantly high, especially with large-scale infrastructures. The need for speed, which is primary goal of HPC, leads to the choice of faster computing units for which the design is mainly guided by processing speed regardless of energy aspects. This is for instance case with GPUs. Training cutting-edge Machine Learning algorithms are handled with large-scale GPU(-enhanced) clusters and their usage is entering into the routine by an increasingly large community (like with the case of ChatGPT), these two facts clearly exacerbate the energy concern.  A complementary fact in this aspect is the “energy as a constraint” standpoint. Embedded systems and mobile platforms are typically battery-powered, thus they run with a fixed amount of energy, which thereby stands as a critical resource. Many applications including AI ones are intended to run on such systems to address common issues, thus the importance of energy efficiency at all levels (supply and consumption). The second reason for the focus on energy is CO$_2$ emission concern with all its consequences. Designing energy-aware solutions is very important and this can be done with several approaches including {\em algorithms design, programs implementations, run-time monitoring tools, compilation, hardware mechanisms, system policies}, and more, beside {\em energy supply} and ways to cap {\em heat dissipation} and {\em CO$_2$ emission}. As HPC is moving on actively through noteworthy processing performance and devices diversity, addressing energy concerns and related aspects is genuinely crucial.

\section*{Acknowledgements}
We wish to warmly thank the Transition Institute 1.5 \footnote{tti.5 - https://the-transition-institute.minesparis.psl.eu/en/} and our home institution Mines Paris-PSL for their support.

\bibliographystyle{IEEEtran}
\bibliography{references}

@incollection{CARDOSO201717,
title = {Chapter 2 - High-performance embedded computing},
booktitle = {Embedded Computing for High Performance},
publisher = {Morgan Kaufmann},
address = {Boston},
pages = {17-56},
year = {2017},
isbn = {978-0-12-804189-5},
doi = {https://doi.org/10.1016/B978-0-12-804189-5.00002-8},
author = {João M.P. Cardoso and José Gabriel F. Coutinho and Pedro C. Diniz}
}

@INPROCEEDINGS{7092602,
  author={Conficoni, Christian and Bartolini, Andrea and Tilli, Andrea and Tecchiolli, Giampietro and Benini, Luca},
  booktitle={Design, Automation and Test in Europe}, 
  title={Energy-aware cooling for hot-water cooled supercomputers}, 
  year={2015},
  volume={},
  number={},
  pages={1353-1358},
  doi={10.7873/DATE.2015.1100}
}

@inproceedings{10.1145/3136014.3136031,
author = {Pereira, Rui and Couto, Marco and Ribeiro, Francisco and Rua, Rui and Cunha, J\'{a}come and Fernandes, Jo\~{a}o Paulo and Saraiva, Jo\~{a}o},
title = {Energy Efficiency across Programming Languages: How Do Energy, Time, and Memory Relate?},
year = {2017},
isbn = {9781450355254},
publisher = {Association for Computing Machinery},
address = {New York, NY, USA},
url = {https://doi.org/10.1145/3136014.3136031},
doi = {10.1145/3136014.3136031},
booktitle = {Proceedings of the 10th ACM SIGPLAN International Conference on Software Language Engineering},
pages = {256–267},
numpages = {12},
location = {Vancouver, BC, Canada},
series = {SLE 2017}
}

@phdthesis{Ikram2018,
author = {Ikram, Muhammad},
year = {2018},
month = {03},
pages = {},
title = {Energy-Efficient GPU-Based High-Performance Computing}
}

@inproceedings{inproceedings_1,
author = {Morán, Marina and Balladini, Javier and Rexachs, Dolores and Rucci, Enzo},
year = {2020},
month = {12},
pages = {1-8},
title = {Towards Management of Energy Consumption in HPC Systems with Fault Tolerance},
doi = {10.1109/ARGENCON49523.2020.9505498}
}

@article{Mantovani2020PerformanceAE,
  title={Performance and energy consumption of HPC workloads on a cluster based on Arm ThunderX2 CPU},
  author={Filippo Mantovani and Marta Garcia-Gasulla and Jos{\'e} Gracia and Esteban Stafford and Fabio Banchelli and Marc Josep-Fabreg{\'o} and Joel Criado-Ledesma and Mathias Nachtmann},
  journal={ArXiv},
  year={2020},
  volume={abs/2007.04868}
}

@inproceedings{Sueur2010DynamicVA,
  title={Dynamic voltage and frequency scaling: the laws of diminishing returns},
  author={Etienne Le Sueur and Gernot Heiser},
  year={2010}
}

@article{article_1,
author = {D'Agostino, Daniele and Merelli, Ivan and Aldinucci, Marco and Cesini, Daniele},
year = {2021},
month = {06},
pages = {1-9},
title = {Hardware and Software Solutions for Energy-Efficient Computing in Scientific Programming},
volume = {2021},
journal = {Scientific Programming},
doi = {10.1155/2021/5514284}
}

@article{article_2,
author = {Beloglazov, Anton and Buyya, Rajkumar and Lee, Young and Zomaya, Albert},
year = {2010},
month = {07},
pages = {},
title = {A Taxonomy and Survey of Energy-Efficient Data Centers and Cloud Computing Systems},
volume = {82},
isbn = {9780123855121},
journal = {Advances in Computers},
doi = {10.1016/B978-0-12-385512-1.00003-7}
}

@Article{en16020890,
AUTHOR = {Kocot, Bartłomiej and Czarnul, Paweł and Proficz, Jerzy},
TITLE = {Energy-Aware Scheduling for High-Performance Computing Systems: A Survey},
JOURNAL = {Energies},
VOLUME = {16},
YEAR = {2023},
NUMBER = {2},
ARTICLE-NUMBER = {890},
ISSN = {1996-1073},
DOI = {10.3390/en16020890}
}

@article{czarnul2019energy,
  title={Energy-aware high-performance computing: survey of state-of-the-art tools, techniques, and environments},
  author={Czarnul, Pawel and Proficz, Jerzy and Krzywaniak, Adam and others},
  journal={Scientific Programming},
  volume={2019},
  year={2019},
  publisher={Hindawi}
}

@inproceedings{maiterth2018energy,
  title={Energy and power aware job scheduling and resource management: Global survey—initial analysis},
  author={Maiterth, Matthias and Koenig, Gregory and Pedretti, Kevin and Jana, Siddhartha and Bates, Natalie and Borghesi, Andrea and Montoya, Dave and Bartolini, Andrea and Puzovic, Milos},
  booktitle={2018 IEEE International Parallel and Distributed Processing Symposium Workshops (IPDPSW)},
  pages={685--693},
  year={2018},
  organization={IEEE}
}

@article{chaudhry2015thermal,
  title={Thermal-aware scheduling in green data centers},
  author={Chaudhry, Muhammad Tayyab and Ling, Teck Chaw and Manzoor, Atif and Hussain, Syed Asad and Kim, Jongwon},
  journal={ACM Computing Surveys (CSUR)},
  volume={47},
  number={3},
  pages={1--48},
  year={2015},
  publisher={ACM New York, NY, USA}
}

@inproceedings{Ramesh2012TechnicalRE,
  title={Technical Report: Energy Management in Embedded Systems Taxonomy},
  author={Umesh Balaji Kothandapani Ramesh and M{\"a}lardalen and S{\'e}verine Sentilles and Ivica Crnkovic},
  year={2012}
}

@article{Oleynik2015RunTimeEO,
  title={Run-Time Exploitation of Application Dynamism for Energy-Efficient Exascale Computing (READEX)},
  author={Yury Oleynik and Michael Gerndt and Joseph Schuchart and Per Gunnar Kjeldsberg and Wolfgang E. Nagel},
  journal={IEEE 18th International Conference on Computational Science and Engineering},
  year={2015},
  pages={347-350}
}

@inproceedings{Vysocky2017MERICAR,
  title={MERIC and RADAR Generator: Tools for Energy Evaluation and Runtime Tuning of HPC Applications},
  author={Ondrej Vysocky and Martin Beseda and Lubom{\'i}r R{\'i}ha and Jan Zapletal and Michael Lysaght and Venkatesh Kannan},
  booktitle={International Conference on High Performance Computing in Science and Engineering},
  year={2017}
}

@article{Schuchart2017TheRF,
  title={The READEX formalism for automatic tuning for energy efficiency},
  author={Joseph Schuchart and Michael Gerndt and Per Gunnar Kjeldsberg and Michael Lysaght and David Hoř{\'a}k and Lubom{\'i}r R{\'i}ha and Andreas Gocht-Zech and Mohammed Sourouri and Madhura Kumaraswamy and Anamika Chowdhury and Magnus Jahre and Kai Diethelm and Othman Bouizi and Umbreen Sabir Mian and Jakub Kruz{\'i}k and Radim Sojka and Martin Beseda and Venkatesh Kannan and Zakaria Bendifallah and Daniel Hackenberg and Wolfgang E. Nagel},
  journal={Computing},
  year={2017},
  volume={99},
  pages={727-745}
}

@INPROCEEDINGS{8323587,
  author={Grant, Ryan E. and Laros, James H. and Levenhagen, Michael and Olivier, Stephen L. and Pedretti, Kevin and Ward, Lee and Younge, Andrew J.},
  booktitle={2017 Eighth International Green and Sustainable Computing Conference (IGSC)}, 
  title={Evaluating energy and power profiling techniques for HPC workloads}, 
  year={2017},
  volume={},
  number={},
  pages={1-8},
  doi={10.1109/IGCC.2017.8323587}}

@article{Jankovi2015MicrocontrollerPC,
  title={Microcontroller power consumption measurement based on PSoC},
  author={Strahinja Jankovi{\'c} and Vujo Drndarevi{\'c}},
  journal={2015 23rd Telecommunications Forum Telfor (TELFOR)},
  year={2015},
  pages={673-676}
}

@article{DJEDIDI2020101805,
    title = {Power profiling and monitoring in embedded systems: A comparative study and a novel methodology based on NARX neural networks},
    journal = {Journal of Systems Architecture},
    volume = {111},
    pages = {101805},
    year = {2020},
    issn = {1383-7621},
    doi = {https://doi.org/10.1016/j.sysarc.2020.101805},
    url = {https://www.sciencedirect.com/science/article/pii/S1383762120300953},
    author = {Oussama Djedidi and Mohand A. Djeziri}
}

@INPROCEEDINGS{likwid2010,  
    author={Treibig, Jan and Hager, Georg and Wellein, Gerhard},  
    booktitle={39th Int. Conf. on Parallel Processing Workshops},   
    title={LIKWID: A Lightweight Performance-Oriented Tool Suite for x86 Multicore Environments},   
    year={2010},  
    volume={},  
    number={},  
    pages={207-216},  
    doi={10.1109/ICPPW.2010.38}
}

@inproceedings{vampir,
  title={The Vampir Performance Analysis Tool-Set},
  author={Andreas Kn{\"u}pfer and Holger Brunst and Jens Doleschal and Matthias Jurenz and Matthias Lieber and Holger Mickler and Matthias S. M{\"u}ller and Wolfgang E. Nagel},
  booktitle={Parallel Tools Workshop},
  year={2008}
}

@INPROCEEDINGS{lo2s,
  author={Ilsche, Thomas and Schöne, Robert and Bielert, Mario and Gocht, Andreas and Hackenberg, Daniel},
  booktitle={IEEE Int. Conf. on Cluster Comp. (CLUSTER)}, 
  title={lo2s — Multi-core System and Application Performance Analysis for Linux}, 
  year={2017},
  volume={},
  number={},
  pages={801-804},
  doi={10.1109/CLUSTER.2017.116}}

@INPROCEEDINGS{lo2s2018,
  author={Ilsche, Thomas and Schöne, Robert and Joram, Philipp and Bielert, Mario and Gocht, Andreas},
  booktitle={2018 IEEE International Parallel and Distributed Processing Symposium Workshops (IPDPSW)}, 
  title={System Monitoring with lo2s: Power and Runtime Impact of C-State Transitions}, 
  year={2018},
  volume={},
  number={},
  pages={712-715},
  doi={10.1109/IPDPSW.2018.00114}}

@INPROCEEDINGS{PAPIenergy,
  author={McCraw, Heike and Ralph, James and Danalis, Anthony and Dongarra, Jack},
  booktitle={2014 IEEE International Conference on Cluster Computing (CLUSTER)}, 
  title={Power monitoring with PAPI for extreme scale architectures and dataflow-based programming models}, 
  year={2014},
  volume={},
  number={},
  pages={385-391},
  doi={10.1109/CLUSTER.2014.6968672}}

@inproceedings{Browne1999PAPIAP,
  title={PAPI: A Portable Interface to Hardware Performance Counters},
  author={Shirley Browne and Christine Deane and George Ho and Philip Mucci},
  year={1999}
}

@Article{en11030620,
AUTHOR = {Boussaada, Zina and Curea, Octavian and Remaci, Ahmed and Camblong, Haritza and Mrabet Bellaaj, Najiba},
TITLE = {A Nonlinear Autoregressive Exogenous (NARX) Neural Network Model for the Prediction of the Daily Direct Solar Radiation},
JOURNAL = {Energies},
VOLUME = {11},
YEAR = {2018},
NUMBER = {3},
ARTICLE-NUMBER = {620},
URL = {https://www.mdpi.com/1996-1073/11/3/620},
ISSN = {1996-1073},
DOI = {10.3390/en11030620}
}

@inproceedings{Schmidt2009ServerRR,
  title={Server Rack Rear Door Heat Exchanger and the New ASHRAE Recommended Environmental Guidelines},
  author={Roger R. Schmidt and Madhusudan K. Iyengar},
  year={2009}
}

@article{PAMBUDI20229509,
title = {The immersion cooling technology: Current and future development in energy saving},
journal = {Alexandria Engineering Journal},
volume = {61},
number = {12},
pages = {9509-9527},
year = {2022},
issn = {1110-0168},
doi = {https://doi.org/10.1016/j.aej.2022.02.059},
url = {https://www.sciencedirect.com/science/article/pii/S1110016822001557},
author = {Nugroho Agung Pambudi and Alfan Sarifudin and Ridho Alfan Firdaus and Desita Kamila Ulfa and Indra Mamad Gandidi and Rahmat Romadhon}
}

@inproceedings{Meyer2013iDataCoolHW,
  title={iDataCool: HPC with Hot-Water Cooling and Energy Reuse},
  author={Nils Meyer and Manfred Ries and Stefan Solbrig and Tilo Wettig},
  booktitle={Inform. Security Conf.},
  year={2013}
}

@INPROCEEDINGS{Nonaka,
  author={Nonaka, Jorji and Hanawa, Toshihiro and Shoji, Fumiyoshi},
  booktitle={2020 IEEE International Conference on Cluster Computing (CLUSTER)}, 
  title={Analysis of Cooling Water Temperature Impact on Computing Performance and Energy Consumption}, 
  year={2020},
  volume={},
  number={},
  pages={169-175},
  doi={10.1109/CLUSTER49012.2020.00027}}

@article{LJUNGDAHL2022117671,
title = {A decision support model for waste heat recovery systems design in Data Center and High-Performance Computing clusters utilizing liquid cooling and Phase Change Materials},
journal = {Applied Thermal Engineering},
volume = {201},
pages = {117671},
year = {2022},
issn = {1359-4311},
doi = {https://doi.org/10.1016/j.applthermaleng.2021.117671},
url = {https://www.sciencedirect.com/science/article/pii/S1359431121010966},
author = {V. Ljungdahl and M. Jradi and C. Veje}
}

@inproceedings{inproceedings_2,
author = {Cabrera, Alberto and Almeida, Francisco and Blanco, Vicente and Nieves, Dagoberto},
year = {2019},
month = {09},
pages = {},
title = {Finding energy efficient hardware configurations under a power cap}
}

@inproceedings{Claude2003,
author = {Tadonki, Claude and Singh, Mitali and Rolim, José and Prasanna, V.},
year = {2003},
month = {09},
pages = {265-268},
title = {Combinatorial Techniques for Memory Power State Scheduling in Energy-Constrained Systems},
volume = {2909},
isbn = {978-3-540-21079-5},
doi = {10.1007/978-3-540-24592-6_24}
}

@inproceedings{leite2014excalibur,
  title={Excalibur: An autonomic cloud architecture for executing parallel applications},
  author={Leite, Alessandro Ferreira and Raiol, Tain{\'a} and Tadonki, Claude and Walter, Maria Emilia MT and Eisenbeis, Christine and de Melo, Alba Cristina Magalhaes Alves},
  booktitle={Proceedings of the 4th International Workshop on Cloud Data and Platforms},
  pages={1--6},
  year={2014}
}

@INPROCEEDINGS{Lin2013,
  author={Lin, Cheng-Yen and Kuan, Chi-Bang and Lee, Jenq Kuen},
  booktitle={2013 42nd International Conference on Parallel Processing}, 
  title={Compilers for Low Power with Design Patterns on Embedded Multicore Systems}, 
  year={2013},
  volume={},
  number={},
  pages={1052-1060},
  doi={10.1109/ICPP.2013.125}}

@phdthesis{DBLP:phd/hal/Stoffel21,
  author    = {Mathieu Stoffel},
  title     = {Static and dynamic approaches for the optimization
               of the energy consumption associated with applications of the High
               Performance Computing {(HPC)} field},
  school    = {Grenoble Alpes University, France},
  year      = {2021}
}

@article{Hackenberg2014HDEEMHD,
    title={HDEEM: High Definition Energy Efficiency Monitoring},
    author={Daniel Hackenberg and Thomas Ilsche and Joseph Schuchart and Robert Sch{\"o}ne and Wolfgang E. Nagel and Marc Simon and Yiannis Georgiou},
    journal={Energy Efficient Supercomp. Workshop},
    year={2014},
  pages={1-10}
}

@INPROCEEDINGS{7307670,
    author={Rashti, Mohammad and Sabin, Gerald and Vansickle, David and Norris, Boyana},
    booktitle={2015 IEEE International Conference on Cluster Computing}, 
    title={WattProf: A Flexible Platform for Fine-Grained HPC Power Profiling}, 
    year={2015},
    volume={},
    number={},
    pages={698-705},
    doi={10.1109/CLUSTER.2015.121}
}

@article{Libri2018DwarfIA,
    title={Dwarf in a Giant: Enabling Scalable, High-Resolution HPC Energy Monitoring for Real-Time Profiling and Analytics},
    author={Antonio Libri and Andrea Bartolini and Luca Benini},
    journal={ArXiv},
    year={2018},
    volume={abs/1806.02698}
}

@article{LEITE20142260,
    title = {A Fine-grained Approach for Power Consumption Analysis and Prediction},
    journal = {Procedia Computer Science (ICCS2014)},
    volume = {29},
    pages = {2260-2271},
    year = {2014},
    issn = {1877-0509},
    doi = {https://doi.org/10.1016/j.procs.2014.05.211},
    author = {Leite, Alessandro and Tadonki, Claude and Eisenbeis, Christine and de Melo, Alba}
}

@inproceedings{claude2004,
author = {Tadonki, Claude and Rolim, Jose},
year = {2004},
month = {03},
pages = {},
title = {An Analytical Model for Energy Minimization},
volume = {3059},
isbn = {978-3-540-22067-1},
doi = {10.1007/978-3-540-24838-5_41}
}

@inproceedings{Mitali2003,
author = {Singh, Mitali and Prasanna, V.},
year = {2003},
month = {05},
pages = {237-252},
title = {Algorithmic Techniques for Memory Energy Reduction},
volume = {2647},
isbn = {978-3-540-40205-3},
doi = {10.1007/3-540-44867-5_20}
}

@book{meyers2005effective,
  title={Effective C++: 55 Specific Ways to Improve Your Programs and Designs},
  author={Meyers, S.},
  isbn={9780132702065},
  series={Addison-Wesley Professional Computing Series},
  url={https://books.google.fr/books?id=Qx5oyB49poYC},
  year={2005},
  publisher={Pearson Education}
}

@ARTICLE{4906989,
    author={Ge, Rong and Feng, Xizhou and Song, Shuaiwen and Chang, Hung-Ching and Li, Dong and Cameron, Kirk W.},
    journal={IEEE Transactions on Parallel and Distributed Systems}, 
    title={PowerPack: Energy Profiling and Analysis of High-Performance Systems and Applications}, 
    year={2010},
    volume={21},
    number={5},
    pages={658-671},
    doi={10.1109/TPDS.2009.76}
}

@article{Schne2021EnergyEA,
  title={Energy Efficiency Aspects of the AMD Zen 2 Architecture},
  author={Robert Sch{\"o}ne and Thomas Ilsche and Mario Bielert and Markus Velten and Markus Schmidl and Daniel Hackenberg},
  journal={IEEE Int. Conference on Cluster Computing (CLUSTER)},
  year={2021},
  pages={562-571}
}

@INPROCEEDINGS{6560768,
  author={Sai, Manoj P. D. and Wang, Kanwen and Yu, Hao},
  booktitle={2013 50th ACM/EDAC/IEEE Design Automation Conference (DAC)}, 
  title={Peak power reduction and workload balancing by space-time multiplexing based demand-supply matching for 3D thousand-core microprocessor}, 
  year={2013},
  volume={},
  number={},
  pages={1-6},
  doi={10.1145/2463209.2488950}}

@article{benoit:hal-01557025,
  TITLE = {{Reducing the energy consumption of large scale computing systems through combined shutdown policies with multiple constraints}},
  AUTHOR = {Benoit, Anne and Lef{\`e}vre, Laurent and Orgerie, Anne-C{\'e}cile and Ra{\"i}s, Issam},
  URL = {https://inria.hal.science/hal-01557025},
  JOURNAL = {{Int. Journal of High Performance Computing Applications}},
  PUBLISHER = {{SAGE Publications}},
  VOLUME = {32},
  NUMBER = {1},
  PAGES = {176-188},
  YEAR = {2018},
  MONTH = Jan,
  DOI = {10.1177/1094342017714530},
  HAL_ID = {hal-01557025},
  HAL_VERSION = {v1},
}

@article{bottleneck_arch,
  author       = {Kaiming He and
                  Xiangyu Zhang and
                  Shaoqing Ren and
                  Jian Sun},
  title        = {Deep Residual Learning for Image Recognition},
  journal      = {CoRR},
  volume       = {abs/1512.03385},
  year         = {2015},
  url          = {http://arxiv.org/abs/1512.03385},
  eprinttype    = {arXiv},
  eprint       = {1512.03385},
  bibsource    = {dblp computer science bibliography, https://dblp.org}
}

@article{Sanjeevi2017,
author = {Sanjeevi, P. and Perumal, Viswanathan},
year = {2017},
month = {01},
pages = {115},
title = {Workload consolidation techniques to optimise energy in cloud: Review},
volume = {10},
journal = {International Journal of Internet Protocol Technology},
doi = {10.1504/IJIPT.2017.085190}
}

@INPROCEEDINGS{9555945,
  author={Schöne, Robert and Schmidl, Markus and Bielert, Mario and Hackenberg, Daniel},
  booktitle={2021 IEEE Int. Conference on Cluster Computing (CLUSTER)}, 
  title={FIRESTARTER 2: Dynamic Code Generation for Processor Stress Tests}, 
  year={2021},
  volume={},
  number={},
  pages={582-590},
  doi={10.1109/Cluster48925.2021.00084}}

@inproceedings{Eastep2017GlobalEO,
  title={Global Extensible Open Power Manager: A Vehicle for HPC Community Collaboration on Co-Designed Energy Management Solutions},
  author={Jonathan M. Eastep and Steve Sylvester and Christopher M. Cantalupo and Brad Geltz and Federico Ardanaz and Asma H. Al-rawi and Kelly Livingston and Fuat Keceli and Matthias Maiterth and Siddhartha Jana},
  booktitle={Information Security Conference},
  year={2017},
  url={https://api.semanticscholar.org/CorpusID:791727}
}

@book{book_1,
author = {Hennessy, John L.  and Patterson, David A. },
year = {2012},
month = {},
pages = {22},
title = {Computer Architecture: A Quantitative Approach (5th ed.)},
volume = {},
isbn = {ISBN 978-0-12-383872-8},
journal = {Elsevier}
}

@inproceedings{Gao2014MachineLA,
  title={Machine Learning Applications for Data Center Optimization},
  author={Jim Gao},
  year={2014},
  url={https://api.semanticscholar.org/CorpusID:64625439}
}

@inproceedings{Belady2008GREENGD,
  title={GREEN GRID DATA CENTER POWER EFFICIENCY METRICS: PUE AND DCIE},
  author={Christian Belady and Andrew Rawson},
  year={2008}
}

@article{Jin2016GreenDC,
  title={Green Data Centers: A Survey, Perspectives and Future Directions},
  author={Xibo Jin and Fa Zhang and Athanasios V. Vasilakos and Zhiyong Liu},
  journal={arXiv:1608.00687},
  year={2016}
}

@INPROCEEDINGS{Coral2015,
  author={Wu, Xingfu and Taylor, Valerie},
  booktitle={2015 Sixth International Green and Sustainable Computing Conference (IGSC)}, 
  title={Power and performance characteristics of CORAL Scalable Science Benchmarks on BlueGene/Q Mira}, 
  year={2015},
  volume={},
  number={},
  pages={1-6},
  doi={10.1109/IGCC.2015.7393681}}

@article{FREITAG2021100340,
title = {The real climate and transformative impact of ICT: A critique of estimates, trends, and regulations},
journal = {Patterns},
volume = {2},
number = {9},
pages = {100340},
year = {2021},
issn = {2666-3899},
doi = {https://doi.org/10.1016/j.patter.2021.100340},
url = {https://www.sciencedirect.com/science/article/pii/S2666389921001884},
author = {Charlotte Freitag and Mike Berners-Lee and Kelly Widdicks and Bran Knowles and Gordon S. Blair and Adrian Friday}
}

@article{Goel2020ASO,
  title={A Survey of Methods for Low-Power Deep Learning and Computer Vision},
  author={Abhinav Goel and Caleb Tung and Yung-Hsiang Lu and George K. Thiruvathukal},
  journal={2020 IEEE 6th World Forum on Internet of Things (WF-IoT)},
  year={2020},
  pages={1-6}
}

@article{distil2021,
author = {Gou, Jianping and Yu, Baosheng and Maybank, Stephen J. and Tao, Dacheng},
title = {Knowledge Distillation: A Survey},
year = {2021},
issue_date = {Jun 2021},
publisher = {Kluwer Academic Publishers},
address = {USA},
volume = {129},
number = {6},
issn = {0920-5691},
url = {https://doi.org/10.1007/s11263-021-01453-z},
doi = {10.1007/s11263-021-01453-z},
journal = {Int. J. Comput. Vision},
month = {jun},
pages = {1789–1819},
numpages = {31}
}

@article{Sanh2019DistilBERTAD,
  title={DistilBERT, a distilled version of BERT: smaller, faster, cheaper and lighter},
  author={Victor Sanh and Lysandre Debut and Julien Chaumond and Thomas Wolf},
  journal={ArXiv:1910.01108},
  year={2019}
}

@INPROCEEDINGS{Frey2022,
  author={Frey, Nathan C. and Zhao, Dan and Axelrod, Simon and Jones, Michael and Bestor, David and Gadepally, Vijay and Gómez-Bombarelli, Rafael and Samsi, Siddharth},
  booktitle={2022 IEEE International Parallel and Distributed Processing Symposium Workshops (IPDPSW)}, 
  title={Energy-aware neural architecture selection and hyperparameter optimization}, 
  year={2022},
  volume={},
  number={},
  pages={732-741},
  doi={10.1109/IPDPSW55747.2022.00125}}

@ARTICLE{Nasrin2022,
  author={Nasrin, Shamma and Shylendra, Ahish and Darabi, Nastaran and Tulabandhula, Theja and Gomes, Wilfred and Chakrabarty, Ankush and Trivedi, Amit Ranjan},
  journal={IEEE Access}, 
  title={ENOS: Energy-Aware Network Operator Search in Deep Neural Networks}, 
  year={2022},
  volume={10},
  number={},
  pages={81447-81457},
  doi={10.1109/ACCESS.2022.3192515}}

@article{wang2019energy,
  title={Energy-Aware Neural Architecture Optimization with Fast Splitting Steepest Descent},
  author={Wang, Dilin and Li, Meng and Wu, Lemeng and Chandra, Vikas and Liu, Qiang},
  journal={arXiv preprint arXiv:1910.03103},
  year={2019}
}

@inproceedings{wu2019splitting,
  title={Splitting steepest descent for growing neural architectures},
  author={Wu, Lemeng and Wang, Dilin and Liu, Qiang},
  booktitle={Advances in Neural Information Processing Systems},
  pages={10655--10665},
  year={2019}
}

@inproceedings{budennyy2023eco2ai,
  title={Eco2ai: carbon emissions tracking of machine learning models as the first step towards sustainable ai},
  author={Budennyy, SA and Lazarev, VD and Zakharenko, NN and Korovin, AN and Plosskaya, OA and Dimitrov, DV and Akhripkin, VS and Pavlov, IV and Oseledets, IV and Barsola, IS and others},
  booktitle={Doklady Mathematics},
  year={2023}
}

@misc{anthony2020carbontracker,
  title={Carbontracker: Tracking and Predicting the Carbon Footprint of Training Deep Learning Models},
  author={Lasse F. Wolff Anthony and Benjamin Kanding and Raghavendra Selvan},
  howpublished={ICML Workshop on Challenges in Deploying and monitoring Machine Learning Systems},
  month={July},
  note={arXiv:2007.03051},
  year={2020}}

@article{Lannelongue2020GreenAQ,
author = {Lannelongue, Loïc and Grealey, Jason and Inouye, Michael},
title = {Green Algorithms: Quantifying the Carbon Footprint of Computation},
journal = {Advanced Science},
volume = {8},
number = {12},
doi = {https://doi.org/10.1002/advs.202100707},
year = {2021}
}

@inproceedings{tado_europar,
  title={The algebraic path problem revisited},
  author={Rajopadhye, Sanjay and Tadonki, Claude and Risset, Tanguy},
  booktitle={Euro-Par’99 Parallel Processing: 5th International Euro-Par Conference Toulouse, France, August 31--September 3, 1999 Proceedings 5},
  pages={698--707},
  year={1999},
  organization={Springer Berlin Heidelberg}
}

@misc{henderson2020systematic,
    title={Towards the Systematic Reporting of the Energy and Carbon Footprints of Machine Learning},
    author={Peter Henderson and Jieru Hu and Joshua Romoff and Emma Brunskill and Dan Jurafsky and Joelle Pineau},
    year={2020},
    eprint={2002.05651},
    archivePrefix={arXiv},
    primaryClass={cs.CY}
}

@INPROCEEDINGS{youssef1,
  author={Zhao, Dan and Frey, Nathan C. and McDonald, Joseph and Hubbell, Matthew and Bestor, David and Jones, Michael and Prout, Andrew and Gadepally, Vijay and Samsi, Siddharth},
  booktitle={2022 IEEE International Parallel and Distributed Processing Symposium Workshops (IPDPSW)}, 
  title={A Green(er) World for A.I.}, 
  year={2022},
  volume={},
  number={},
  pages={742-750},
  doi={10.1109/IPDPSW55747.2022.00126}}

@InProceedings{youssef2,
    author="Caspart, Ren{\'e} and Ziegler, Sebastian and Weyrauch, Arvid and Obermaier, Holger and Raffeiner, Simon and Schuhmacher, Leon Pascal and Scholtyssek, Jan and Trofimova, Darya and Nolden, Marco and Reinartz, Ines and Isensee, Fabian and G{\"o}tz, Markus and Debus, Charlotte",
    editor="Anzt, Hartwig and Bienz, Amanda and Luszczek, Piotr and Baboulin, Marc",
    title="Precise Energy Consumption Measurements of Heterogeneous Artificial Intelligence Workloads",
    booktitle="High Performance Computing. ISC High Performance 2022 International Workshops",
    year="2022",
    publisher="Springer International Publishing",
    address="Cham",
    pages="108--121"
}

@article{youssef3,
  title={A Survey of Quantization Methods for Efficient Neural Network Inference},
  author={Amir Gholami and Sehoon Kim and Zhen Dong and Zhewei Yao and Michael W. Mahoney and Kurt Keutzer},
  journal={ArXiv},
  year={2021},
  volume={abs/2103.13630}
}

@misc{youssef4,
      title={Understanding the Energy Consumption of HPC Scale Artificial Intelligence}, 
      author={Danilo Carastan dos Santos},
      year={2022},
      eprint={2212.00582},
      archivePrefix={arXiv},
      primaryClass={cs.DC}
}

@article{Wu2021SustainableAE,
  title={Sustainable AI: Environmental Implications, Challenges and Opportunities},
  author={Carole-Jean Wu and Ramya Raghavendra and Udit Gupta and Bilge Acun and Newsha Ardalani and Kiwan Maeng and Gloria Chang and Fiona Aga Behram and James Huang and Charles Bai and Michael K. Gschwind and Anurag Gupta and Myle Ott and Anastasia Melnikov and Salvatore Candido and David Brooks and Geeta Chauhan and Benjamin Lee and Hsien-Hsin S. Lee and Bugra Akyildiz and Maximilian Balandat and Joe Spisak and Ravi Kumar Jain and Michael G. Rabbat and Kim M. Hazelwood},
  journal={arXiv:2111.00364},
  year={2021}
}

@inproceedings{inproceedings_3,
author = {Strubell, Emma and Ganesh, Ananya and Mccallum, Andrew},
year = {2019},
month = {01},
pages = {3645-3650},
title = {Energy and Policy Considerations for Deep Learning in NLP},
doi = {10.18653/v1/P19-1355}
}

@ARTICLE{Patterson2022,
  author={Patterson, David and Gonzalez, Joseph and Hölzle, Urs and Le, Quoc and Liang, Chen and Munguia, Lluis-Miquel and Rothchild, Daniel and So, David R. and Texier, Maud and Dean, Jeff},
  journal={Computer}, 
  title={The Carbon Footprint of Machine Learning Training Will Plateau, Then Shrink}, 
  year={2022},
  volume={55},
  number={7},
  pages={18-28},
  doi={10.1109/MC.2022.3148714}
}

@article{Yang2016DesigningEC,
  title={Designing Energy-Efficient Convolutional Neural Networks Using Energy-Aware Pruning},
  author={Tien-Ju Yang and Yu-hsin Chen and Vivienne Sze},
  journal={2017 IEEE Conference on Computer Vision and Pattern Recognition (CVPR)},
  year={2016},
  pages={6071-6079}
}

@inproceedings{DBLP:conf/iscc/VaddinaLO21,
  author    = {Kameswar Rao Vaddina and
               Laurent Lef{\`{e}}vre and
               Anne{-}C{\'{e}}cile Orgerie},
  title     = {Experimental Workflow for Energy and Temperature Profiling on {HPC}
               Systems},
  booktitle = {{IEEE} Symposium on Computers and Communications, {ISCC} 2021, Athens,
               Greece, September 5-8, 2021},
  pages     = {1--7},
  publisher = {{IEEE}},
  year      = {2021},
  url       = {https://doi.org/10.1109/ISCC53001.2021.9631413},
  doi       = {10.1109/ISCC53001.2021.9631413},
  bibsource = {dblp computer science bibliography, https://dblp.org}
}

@misc{qiu2022look,
      title={A first look into the carbon footprint of federated learning}, 
      author={Xinchi Qiu and Titouan Parcollet and Javier Fernandez-Marques and Pedro Porto Buarque de Gusmao and Yan Gao and Daniel J. Beutel and Taner Topal and Akhil Mathur and Nicholas D. Lane},
      year={2022},
      eprint={2102.07627},
      archivePrefix={arXiv},
      primaryClass={cs.LG}
}

@article{Luccioni2022EstimatingTC,
  title={Estimating the Carbon Footprint of BLOOM, a 176B Parameter Language Model},
  author={Alexandra Sasha Luccioni and Sylvain Viguier and Anne-Laure Ligozat},
  journal={ArXiv},
  year={2022},
  volume={abs/2211.02001}
}

@article{Patterson2021CarbonEA,
  title={Carbon Emissions and Large Neural Network Training},
  author={David A. Patterson and Joseph Gonzalez and Quoc V. Le and Chen Liang and Llu{\'i}s-Miquel Mungu{\'i}a and Daniel Rothchild and David R. So and Maud Texier and Jeff Dean},
  journal={ArXiv},
  year={2021},
  volume={abs/2104.10350}
}

@article{lacoste2019quantifying,
  title={Quantifying the Carbon Emissions of Machine Learning},
  author={Lacoste, Alexandre and Luccioni, Alexandra and Schmidt, Victor and Dandres, Thomas},
  journal={arXiv:1910.09700},
  year={2019}
}

@phdthesis{tadHDR,
  TITLE = {High Performance Computing as a Combination of Machines and Methods and Programming},
  AUTHOR = {Tadonki, Claude},
  URL = {https://theses.hal.science/tel-00832930},
  SCHOOL = {{Universit{\'e} Paris Sud - Paris XI}},
  YEAR = {2013},
  MONTH = May,
  TYPE = {Habilitation {\`a} diriger des recherches},
  HAL_ID = {tel-00832930},
  HAL_VERSION = {v1},
}

@article{Hosseinabady2018DynamicEM,
  title={Dynamic Energy Management of FPGA Accelerators in Embedded Systems},
  author={Mohammad Hosseinabady and Jos{\'e} Luis N{\'u}{\~n}ez-Y{\'a}{\~n}ez},
  journal={ACM Transactions on Embedded Computing Systems (TECS)},
  year={2018},
  volume={17},
  pages={1 - 26}
}

@INPROCEEDINGS{9586224,
  author={Pandey, Pramesh and Gundi, Noel Daniel and Chakraborty, Koushik and Roy, Sanghamitra},
  booktitle={2021 58th ACM/IEEE Design Automation Conference (DAC)}, 
  title={UPTPU: Improving Energy Efficiency of a Tensor Processing Unit through Underutilization Based Power-Gating}, 
  year={2021},
  volume={},
  number={},
  pages={325-330},
  doi={10.1109/DAC18074.2021.9586224}}

@INPROCEEDINGS{fpga2022,
  author={Boku, Taisuke},
  booktitle={2022 International Symposium on VLSI Design, Automation and Test (VLSI-DAT)}, 
  title={How FPGA can contribute to HPC ?}, 
  year={2022},
  volume={},
  number={},
  doi={10.1109/VLSI-DAT54769.2022.9768098}}

@ARTICLE{9118980,
  author={Jafari-Nodoushan, Mostafa and Safaei, Bardia and Ejlali, Alireza and Chen, Jian-Jia},
  journal={IEEE Trans. on Circuits and Systems I}, 
  title={Leakage-Aware Battery Lifetime Analysis Using the Calculus of Variations}, 
  year={2020},
  volume={67},
  number={12},
  pages={4829-4841},
  doi={10.1109/TCSI.2020.3001064}
}

@inproceedings{zeus-nsdi23,
    title     = {Zeus: Understanding and Optimizing {GPU} Energy Consumption of {DNN} Training},
    author    = {Jie You and Jae-Won Chung and Mosharaf Chowdhury},
    booktitle = {USENIX NSDI},
    year      = {2023}
}

@misc{Frontier2022,
  author= {ORNL},
  year  = {2022},
  title = {FRONTIER DIRECTION OF DISCOVERY},
  url = {https://www.olcf.ornl.gov/frontier/},
  note = {Accessed: 2023-05-04}
}

@misc{top500,
  author= {Erich, Strohmaier and Jack, Dongarra and Horst , Simon and Martin, Meuer},
  year  = {2022},
  title = {TOP500},
  url = {https://www.top500.org/lists/top500/2022/11/}}

@misc{green500,
  author= {Kirk, W., Cameron},
  year  = {2022},
  title = {GREEN500},
  url = {https://www.top500.org/lists/green500/},
  note = {Accessed: 2023-05-04}
}

@misc{acp2011,
  author= {Scott, Huck},
  year  = {2011},
  title = {Measuring Processor Power TDP vs ACP : withe paper},
  url = {https://www.intel.com/content/dam/doc/white-paper/resources-xeon-measuring-processor-power-paper.pdf},
  note = {Accessed: 2023-04-21}
}

@misc{acp2022,
  author= {Arne, Tarara},
  year  = {2022},
  title = {TDP AND ACP FOR ENERGY ESTIMATION IN PROCESSORS},
  url = {https://www.green-coding.berlin/blog/tdp-and-acp/},
  note = {Accessed: 2023-04-21}
}

@misc{SWaP2013,
  author= {David, Greenhill},
  year  = {2013},
  title = {SWaP Space Watts and Power},
  url = {https://www.energystar.gov/ia/products/downloads/Greenhill\_Pres.pdf},
  note = {Accessed: 2023-04-21}
}

@misc{uptime2021,
  author= {Daniel, Bizo and Rhonda, Ascierto and Andy, Lawrence and Jacqueline, Davis},
  year  = {2021},
  title = {2021 Data Center Industry Survey Results},
  url = {https://uptimeinstitute.com/2021-data-center-industry-survey-results},
  note = {Accessed: 2023-04-21}
}

@misc{uptime2022,
  author= {Jacqueline, Davis and Daniel, Bizo and Andy, Lawrence and Owen, Rogers and Max, Smolaks},
  year  = {2022},
  title = {2022 Data Center Industry Survey Results},
  url = {https://uptimeinstitute.com/resources/research-and-reports/uptime-institute-global-data-center-survey-results-2022},
  note = {Accessed: 2023-04-21}
}

@misc{astra,
  author= {Wikichip.org},
  year  = {2020},
  title = {Astra - Supercomputers},
  url = {https://en.wikichip.org/wiki/supercomputers/astra},
  note = {Accessed: 2023-05-16}
}

@misc{Lakefield,
  author= {Intel},
  year  = {2019},
  title = {HeadlineLakefield: Hybrid CPU with Foveros Technology},
  url = {https://www.intel.com/content/www/us/en/newsroom/resources/lakefield.html},
  note = {Accessed: 2023-05-16}
}

@misc{acpi,
  author= {UEFI},
  year  = {2021},
  title = {Advanced Configuration and Power Interface (ACPI) Specification, January 2021},
  url = {https://uefi.org/htmlspecs/ACPI_Spec_6_4_html/},
  note = {Accessed: 2023-05-16}
}

@misc{devicetree,
  author= {Linaro},
  year  = {2022},
  title = {The Devicetree Specification},
  url = {https://www.devicetree.org/specifications/},
  note = {Accessed: 2023-05-16}
}

@misc{nvidia2016,
  author= {NVIDIA},
  year  = {2016},
  title = {nvidia-smi - NVIDIA System Management Interface program},
  url = {https://developer.download.nvidia.com/compute/DCGM/docs/nvidia-smi-367.38.pdf},
  note = {Accessed: 2023-04-21}
}

@misc{ROCm-SMI,
  author= {AMD},
  year  = {2014},
  title = {ROCm System Management},
  url = {https://sep5.readthedocs.io/en/latest/ROCm_System_Managment/ROCm-System-Managment.html},
  note = {Accessed: 2023-04-21}
}

@misc{nvidiaGPU,
  author= {NVIDIA},
  year  = {2022},
  title = {NVIDIA H100 Tensor Core GPU},
  url = {https://www.nvidia.com/en-us/data-center/h100/},
  note = {Accessed: 2023-05-10}
}

@misc{amd2021,
  author= {AMD},
  year  = {2021},
  title = {AMD INSTINCT MI200 SERIES ACCELERATOR},
  url = {https://www.amd.com/system/files/documents/amd-instinct-mi200-datasheet.pdf},
  note = {Accessed: 2023-04-21}
}

@misc{IntelGPU2022,
  author= {Intel},
  year  = {2022},
  title = {Intel Data Center GPU Max 1550},
  url = {https://ark.intel.com/content/www/us/en/ark/products/232873/intel-data-center-gpu-max-1550.html},
  note = {Accessed: 2023-05-10}
}

@misc{tpu2023,
  author= {Google},
  year  = {2023},
  title = {system architecture tpu v4},
  url = {https://cloud.google.com/tpu/docs/system-architecture-tpu-vm},
  note = {Accessed: 2023-04-21}
}

@misc{eembc2014,
  author= {EEMBC},
  year  = {2014},
  title = {An EEMBC Benchmark},
  url = {https://www.eembc.org/ulpmark/ulp-cp/},
  note = {Accessed: 2023-04-21}
}

@misc{coremark,
  author= {EEMBC},
  year  = {2023},
  title = {About CoreMark-PRO},
  url = {https://www.eembc.org/coremark-pro/},
  note = {Accessed: 2023-05-23}
}

@misc{Drwattson,
  author= {UpbeatLabs},
  year  = {2023},
  title = {Dr. Wattson Energy Monitoring Module for Arduino, Raspberry Pi and other Maker-Friendly Microcontrollers},
  howpublished = {\url{https://www.upbeatlabs.com/wattson/}},
  note = {Accessed: 2023-05-23}
}

@misc{RAPL2022,
  author= {Intel, Corporation},
  year  = {2022},
  title = {Running Average Power Limit Energy Reporting / CVE-2020-8694 , CVE-2020-8695 / INTEL-SA-00389},
  url = {https://www.intel.com/content/www/us/en/developer/articles/technical/software-security-guidance/advisory-guidance/running-average-power-limit-energy-reporting.html},
  note = {Accessed: 2023-04-21}
}

@misc{perftools,
  author= {$\cdots$},
  year  = {2023},
  title = {Energy estimates},
  url = {https://firefox-source-docs.mozilla.org/performance/perf.html},
  note = {Accessed: 2023-05-26}
}

@misc{powergadget,
  author= {Intel},
  year  = {2019},
  title = {Intel Power Gadget},
  url = {https://www.intel.com/content/www/us/en/developer/articles/tool/power-gadget.html},
  note = {Accessed: 2023-05-26}
}

@misc{powerstat,
  author= {Colin, Ian, King},
  year  = {2021},
  title = {powerstat - a tool to measure power consumption},
  url = {https://manpages.ubuntu.com/manpages/bionic/man8/powerstat.8.html},
  note = {Accessed: 2023-05-26}
}

@misc{pyjoules,
  author= {INRIA-Lille},
  year  = {2019},
  title = {Welcome to pyJoules’s documentation!},
  howpublished = {\url{https://pyjoules.readthedocs.io/en/latest/}},
  note = {Accessed: 2023-05-26}
}

@misc{powertop,
  author= {Intel},
  year  = {2020},
  title = {PowerTOP},
  url = {https://github.com/fenrus75/powertop},
  note = {Accessed: 2023-05-26}
}

@misc{tx2mon,
  author= {Marvell},
  year  = {2019},
  title = {tx2mon},
  url = {https://github.com/jchandra-cavm/tx2mon},
  note = {Accessed: 2023-05-16}
}

@misc{grid5000,
  author= {Grid500},
  year  = {2023},
  title = {Grid5000:Home},
  url = {https://www.grid5000.fr/w/Grid5000:Home},
  note = {Accessed: 2023-04-21}
}

@misc{anne2021,
  author= {Anne-Cécile, Orgerie},
  year  = {2021},
  title = {Measuring the Energy Consumption of HPC Systems},
  url = {https://ecoinfo.cnrs.fr/wp-content/uploads/2021/12/ORAP-2021-Orgerie.pdf},
  note = {Accessed: 2023-04-21}
}

@misc{lenovo2022,
  author= {Lenovo},
  year  = {2022},
  title = {Optimizing Power and Energy in HPC Data Centers with Energy Aware Runtime},
  url = {https://lenovopress.lenovo.com/lp1646.pdf},
  note = {Accessed: 2023-04-21}
}

@misc{torwald2021,
  author= {Alberto, Romero},
  year  = {2021},
  title = {Meet M6 - 10 Trillion Parameters at 1\% GPT-3’s Energy Cost},
  url = {https://towardsdatascience.com/meet-m6-10-trillion-parameters-at-1-gpt-3s-energy-cost-997092cbe5e8},
  note = {Accessed: 2023-04-21}
}

@misc{openAI2018,
  author= {Dario, Amodei and Danny, Hernandez},
  year  = {2018},
  title = {AI and compute},
  url = {https://openai.com/research/ai-and-compute},
  note = {Accessed: 2023-04-21}
}

@misc{ncs22022,
  author= {Intel},
  year  = {2022},
  title = {Intel Neural Compute Stick 2 (Intel NCS2)},
  url = {https://www.intel.com/content/www/us/en/developer/articles/tool/neural-compute-stick.html},
  note = {Accessed: 2023-04-21}
}

@misc{ncs22017,
  author= {Paul, Alcorn},
  year  = {2017},
  title = {Intel Unveils Movidius Myriad X Vision Processing Unit},
  url = {https://www.tomshardware.com/news/intel-movidius-vpu-ai-inference,35327.html},
  note = {Accessed: 2023-04-21}
}

@misc{devboard2020,
  author= {GoogleLLC},
  year  = {2020},
  title = {Edge TPU performance benchmarks},
  url = {https://coral.ai/docs/edgetpu/benchmarks/}
}

@misc{coral2020,
  author= {GoogleLLC},
  year  = {2020},
  title = {Dev Board datasheet},
  url = {https://coral.ai/docs/dev-board/datasheet/},
  note = {Accessed: 2023-05-05}
}

@misc{RaspberryPi,
  author= {RaspberryPi},
  year  = {2020},
  title = {Raspberry Pi 4 model B},
  url = {https://www.raspberrypi.com/products/raspberry-pi-4-model-b/},
  note = {Accessed: 2023-05-11}
}

@misc{JetsonNano,
  author= {Nvidia},
  year  = {2020},
  title = {NVIDIA Jetson Nano module Product details},
  url = {https://www.seeedstudio.com/NVIDIAr-Jetson-Nanotm-Developer-Kit-p-2916.html},
  note = {Accessed: 2023-05-11}
}

@misc{JetsonAGX,
  author= {Nvidia},
  year  = {2023},
  title = {NVIDIA Jetson AGX Orin for Robotics and Edge AI Applications},
  url = {https://www.nvidia.com/en-us/lp/embedded-computing/robotics-edge-ai-tech-brief/},
  note = {Accessed: 2023-05-11}
}

@misc{Arduino,
  author= {$\cdots$},
  year  = {2023},
  title = {Arduino Portenta H7 product Overview},
  url = {https://store.arduino.cc/products/portenta-h7},
  note = {Accessed: 2023-05-11}
}

@misc{amd_vs_intel_cpu,
  author= {$\cdots$},
  year  = {2023},
  title = {AMD EPYC Energy Efficiency},
howpublished = "\url{https://www.amd.com/en/campaigns/epyc-energy-efficiency}",
  note = {Accessed: 2023-05-16}
}

@misc{price2022,
  author= {GlobalPetrolPrice.com},
  year  = {2022},
  title = {Global Electricity prices September 2022},
  url = {https://www.globalpetrolprices.com/electricity_prices/},
  note = {Accessed: 2023-05-10}
}

@misc{elec_carbon,
  author= {ourworldindata.org},
  year  = {2023},
  title = {Electricity Mix : Carbon intensity of electricity},
  url = {https://ourworldindata.org/electricity-mix},
  note = {Accessed: 2023-05-11}
}

@misc{tx2,
  author= {Marvell},
  year  = {2019},
  title = {Manufacturing Applications on Marvell ThunderX2: Withe paper},
  url = {https://www.marvell.com/content/dam/marvell/en/products/assets/server-processors/thunderx2-arm-processors/pdf/},
  note = {Accessed: 2023-05-26}
}

@misc{Ludvigsen2022,
  author= {Kasper Groes Albin Ludvigsen},
  year  = {2022},
  title = {How to estimate and reduce the carbon footprint of machine learning models},
  url = {https://towardsdatascience.com/how-to-estimate-and-reduce-the-carbon-footprint-\of-machine-learning-models-49f24510880},
  note = {Accessed: 2023-05-26}
}

@misc{codecarbon,
  author= {GAMMA BCG},
  year  = {2020},
  title = {CodeCarbon},
  url = {https://mlco2.github.io/codecarbon/index.html},
  note = {Accessed: 2023-05-26}
}

@misc{Tracarbon,
  author= {Florian Valeye},
  year  = {2022},
  title = {Tracarbon — Track your device’s carbon footprint},
  url = {https://medium.com/@florian.valeye/tracarbon-track-your-devices-carbon-footprint-fb051fcc9009},
  note = {Accessed: 2023-07-12}
}

@misc{cooling,
  author= {scientific-computing.com},
  year  = {2019},
  title = {Cooling technology options for HPC},
  url = {https://www.scientific-computing.com/feature/cooling-technology-options-hpc},
  note = {Accessed: 2023-06-12}
}

@misc{RDHX-benef,
  author= {AKCP},
  year  = {2021},
  title = {Rear-Door Heat Exchanger for High-density Data Center},
  url = {https://www.akcp.com/articles/rear-door-heat-exchanger-for-high-density-data-center/},
  note = {Accessed: 2023-06-25}
}

@misc{liquid_cooling,
  author= {Systems XENON},
  year  = {2023},
  title = {Liquid Cooling: Exceeding the Limits of Air Cooling to Unlock Greater Potential in HPC},
  url = {https://xenon.com.au/white-papers/liquid-cooling-exceeding-the-limits-of-air-cooling-to-unlock-\greater-potential-in-hpc/},
  note = {Accessed: 2023-06-12}
}

@misc{elec:fugaku, 
year  = {2020},
author= {Arm},
title = {How supercomputer performance and power efficiency can coexist},
url = {https://armkeil.blob.core.windows.net/developer/Files/pdf/case-study/arm-riken-fugaku-supercomputer.pdf},
note = {Accessed: 2023-06-12}
}

@misc{elec_houses, 
year  = {2022},
author= {$\cdots$},
title = {Supercomputers are faster and more powerful — but need to be more energy-efficient},
url ={https://www.businessinsider.in/tech/enterprise/news/},
note = {Accessed: 2023-06-12}
}

\newpage

\section{Biography Section}
% your own custom macro containing the $\backslash${\tt{includegraphics}} command to make things
% To be detailed at the final steps in case of acceptance of the paper. \\
% {\bf Roblex Nana}, PhD student at Mines Paris - PSL University.\\
% {\bf Claude Tadonki}, researcher at Mines Paris - PSL University.\\
% {\bf Petr Dokladal}, researcher at Mines Paris - PSL University.\\
% {\bf Youssef Mesri }, researcher at Mines Paris - PSL University.\\

\vspace{-33pt}

\begin{IEEEbiographynophoto}{Roblex Nana}
Roblex Nana is a doctoral student in High Performance Computing and Artificial Intelligence at Mines Paris - PSL University, studying under Professor Claude Tadonki. Before coming to Paris, he completed a Master degree in computer science in 2022 at University of Yaounde I, Cameroon. Before that, he received a BSc in computer science from the same university in 2019. The topic of his PhD research is "Carbon aware High Performance Artificial Intelligence". 
\end{IEEEbiographynophoto}

\begin{IEEEbiographynophoto}{Claude Tadonki}
Claude Tadonki (M) is a senior researcher and lecturer at the MINES ParisTech Institute (Paris/France) since 2011. He holds a PhD and an HDR in computer science from University of Rennes and from Paris-Sud University respectively . After six years of cutting-edge research in operational research and theoretical computer science at the University of Geneva, he relocated to France to work for EMBL, University of Paris-Sud, LAL-CNRS and then MINES ParisTech. His main research topics included High Performance Computing, Parallel Computing, Operational Research, Matrix Computation, Combinatorial Algorithm and Complexity, Mathematical Programming, Scientific and Technical Programming, Automatic Code Transformations. Claude Tadonki has worked at several laboratories and universities, has initiated various scientific projects and national/international collaborations, has given significant number of CS courses in different contexts including industries. He is an active member of well established scientific corporations and reviewer of high-impact international journals and top-rank conferences. He has published numerous papers in journals and international conferences.He is very active in international collaborations and has co-organized several HPC conferences and forums.
\end{IEEEbiographynophoto}

\begin{IEEEbiographynophoto}{Petr Dokladal}
Petr Dokladal has graduated from the Technical University in Brno, Czech Republic, in 1994 as a Telecommunication Engineer. He obtained his Ph.D. degree in January 2000 from the University of Marne la Valle, France, in General Computer Sciences, specialization Image Processing. He has spent ten months, participating in the COMOBIO project, with the Image and Signal Processing Department at the Ecole des Télécommunications in Paris. In November 2000 he joined the Centre of Mathematical Morphology at the School of Mines in Paris as a research engineer. His interests include medical imaging, image segmentation, object tracking and pattern recognition.
\end{IEEEbiographynophoto}

\begin{IEEEbiographynophoto}{Youssef Mesri}
Youssef Mesri, Associate Professor at Mines ParisTech, joined CEMEF -Center of Material Forming- in January 2015 to develop research and teaching activities in computational high performance mechanics. He is director of the Master program in High Performance Computing and Data Science and deputy head of Computing and Fluids research group. Ex-engineer of IFP Energies Nouvelles, specializing in high-performance numerical methods, Youssef Mesri is one of the developers of the Arcane computing platform to model and analyze fluid mechanics phenomena. Initially developed for nuclear applications (CEA), Arcane is now enriched with new applications, such as multi-phased multi-species flows in geosciences, enabling it to widen the scope of its potential use. He is reviewer of several international journals, including the Computer Methods in Applied Mechanics and Engineering, the Journal of Computational Physics, and Computers and Fluids. His research is about efficient algorithms for the simulation of incompressible/compressible materials using high performance computing methods designed for unstructured meshes.
\end{IEEEbiographynophoto}

%\bf{If you include a photo:}\vspace{-33pt}
%\begin{IEEEbiography}[{\includegraphics[width=1in,height=1.25in,clip,keepaspectratio]{fig1}}]{Michael Shell}
%Use $\backslash${\tt{begin\{IEEEbiography\}}} and then for the 1st argument use $\backslash${\tt{includegraphics}} to declare and link the author photo.
%Use the author name as the 3rd argument followed by the biography text.
%\end{IEEEbiography}


% \begin{IEEEbiographynophoto}{Roblex Nana}
% , PhD student at Mines Paris - PSL University.
% \begin{IEEEbiographynophoto}{Claude Tadonki}
% , researcher at Mines Paris - PSL University.

\vfill

\end{document}